\documentclass[12pt,a4paper]{article}
\usepackage{graphicx}
\usepackage{amssymb}
\usepackage{amsmath}
\usepackage{bm}
\usepackage{color}
\usepackage{theorem}
\usepackage{subfigure}
\usepackage{amsfonts}
\usepackage{dsfont}
\usepackage{nicefrac}
\usepackage[dvipsnames]{xcolor}
\usepackage[export]{adjustbox}
\usepackage{comment}
\usepackage{float}
\usepackage{xfrac}
\usepackage{slashed}
\allowdisplaybreaks

\usepackage{tikz}
\usetikzlibrary{arrows,shapes}
\usetikzlibrary{trees}
\usetikzlibrary{matrix,arrows} 				
\usetikzlibrary{calc} 
\usetikzlibrary{positioning}				
\usetikzlibrary{calc,through}				
\usetikzlibrary{decorations.pathreplacing}  
\usepackage[tikz]{bclogo} 					
\usepackage{pgffor}							

\usetikzlibrary{decorations.pathmorphing}	
\usetikzlibrary{decorations.markings}
\usetikzlibrary{intersections}				
\tikzset{
    vector/.style={decorate, decoration={snake}, draw},
    graviton/.style={decorate, decoration={snake,amplitude=1.5pt}, draw},
    fermion/.style={postaction={decorate},
        decoration={markings,mark=at position .55 with {\arrow{>}}}},
    fermionbar/.style={draw, postaction={decorate},
        decoration={markings,mark=at position .55 with {\arrow{<}}}},
    fermionnoarrow/.style={},
    gluon/.style={decorate,
        decoration={coil,amplitude=4pt, segment length=5pt}},
    scalar/.style={dashed, postaction={decorate},
        decoration={markings,mark=at position .55 with {\arrow{>}}}},
    scalarbar/.style={dashed, postaction={decorate},
        decoration={markings,mark=at position .55 with {\arrow{<}}}},
    scalarnoarrow/.style={dashed,draw},
	provector/.style={decorate, decoration={snake,amplitude=2.5pt}, draw},
	antivector/.style={decorate, decoration={snake,amplitude=-2.5pt}, draw},
	    electron/.style={draw=black, postaction={decorate},
        decoration={markings,mark=at position .55 with {\arrow[draw=black]{>}}}},
	bigvector/.style={decorate, decoration={snake,amplitude=4pt}, draw},
	vectorscalar/.style={loosely dotted,draw=black, postaction={decorate}},
}

\usepackage[sort&compress,numbers, merge]{natbib}

\definecolor{Orange}{cmyk}{0,0.61,0.87,0}
\definecolor{JungleGreen}{cmyk}{0.99,0,0.52,0}
\definecolor{OliveGreen}{cmyk}{0.64,0,0.95,0.40}
\definecolor{Brown}{cmyk}{0,0.81,1,0.60}
\definecolor{RoyalBlue}{cmyk}{0.71,0.53,0,0.12}

\newcommand{\be}{\begin{equation}}
\newcommand{\ee}{\end{equation}}
\newcommand{\bea}{\begin{eqnarray}}
\newcommand{\eea}{\end{eqnarray}}
\newcommand{\eq}[1]{Eq.~(\ref{#1})}
\newcommand{\la}{\langle}
\newcommand{\ra}{\rangle}

\DeclareRobustCommand{\Eq}[1]{Eq.~(\ref{#1})}

\usepackage[colorlinks=True, citecolor=blue, linkcolor=blue, urlcolor=black,linktocpage=true]{hyperref}

\allowdisplaybreaks[1]

\newcommand{\psitil}[0]{\bar{\psi}}

\begin{document}

\begin{center}
\hfill TUM-HEP-1410/22 \\
\hfill BONN-TH-2023-01

\vspace{2.0cm}
{\LARGE \sc
{Two-Loop Infrared Renormalization\\
\vspace{.2cm}with On-shell Methods}
}

\vspace{1.0cm}
{\small 
Pietro Baratella$^a$, Sara Maggio$^b$, Michael Stadlbauer$^{a,c}$ and Tobias Theil$^a$
}

\vspace{0.7cm}
{\it\footnotesize
$^a$Technische Universit\"{a}t M\"{u}nchen, Physik-Department, 85748 Garching, Germany\\
}

{\it\footnotesize
$^b$Bethe Center for Theoretical Physics, Universit\"at Bonn, D-53115, Germany\\
}

{\it\footnotesize
$^c$Max Planck Institute for Physics, F\"ohringer Ring 6, 80805 M\"unchen, Germany\\
}

\vspace{0.9cm}

\abstract{\vspace{.2cm}\noindent Within the framework proposed by Caron-Huot and Wilhelm, we give a recipe for computing infrared anomalous dimensions purely on-shell, efficiently up to two loops in any massless theory. After introducing the general formalism and reviewing the one-loop recipe, we extract a practical formula that relates two-loop infrared anomalous dimensions to certain two- and three-particle phase space integrals with tree-level form factors of conserved operators. We finally provide several examples of the use of the two-loop formula and comment on some of its formal aspects, especially the cancellation of `one-loop squared' spurious terms. The present version of the paper is augmented with a detailed treatment of the structure of infrared divergences in massless theories of scalars and fermions up to two loops. In the calculation we encounter divergent phase space integrals and show in detail how these cancel among each other as required by the finiteness of the anomalous dimension. As a non-trivial check of the method, we also perform the computation with a standard diagrammatic approach, finding perfect agreement.}

\newpage

{
  \hypersetup{linkcolor=blue}
  \tableofcontents
}

\end{center}
\setcounter{footnote}{0}

\section{Introduction}
In reference \cite{Caron-Huot:2016cwu}, the following formula was proposed (see also \cite{EliasMiro:2020tdv} for a very transparent derivation)
\be\label{chw}
e^{-i\pi D}F^*_{\cal O}=S F^*_{\cal O}\,,
\ee
relating the action of the dilation operator $D$ on some renormalized form factor of an operator $\cal O$, called $F_{\cal O}$, to that of the $S$ matrix on the same $F_{\cal O}$.

Despite its indisputable beauty, the formula, as it is, is hard to put into use. The literature that followed this path has mainly focussed on its one-loop implementation \cite{EliasMiro:2020tdv,Baratella:2020lzz}, and on some special two-loop examples which, for a reason or another, are free from infrared (IR) divergences \cite{EliasMiro:2020tdv,Bern:2020ikv,EliasMiro:2021jgu}.
The purpose of this note is to make the first steps into the implementation of \eq{chw} at two loops in cases where IR divergences are present. The key to this will be
\begin{itemize}
\item[(i)] the use of form factors where $\cal O$ is a conserved operator and
\item[(ii)] the use of general results about the structure of IR divergences, especially factorization.
\end{itemize}

In this work we will mainly focus on massless $\lambda\varphi^4$ theory and Yukawa theory, whose infrared divergences are purely collinear and simpler to treat, and only make a few remarks on maximally supersymmetric Yang-Mills. However, insight into these examples will instruct us on the more general scenario.

Following the notation of \cite{EliasMiro:2020tdv}, where $F_{\cal O}({n})\equiv\, _{\rm out}\langle {n}|{\cal O}(0) |0\rangle$ and $S_{n,m}\equiv\, _{\rm out}\langle {n}| {m}\rangle_{\rm in}$, we can rewrite \eq{chw} slightly more explicitly as
\be\label{master0}
e^{-i\pi D}F^*_{\cal O}({n})=\sum_{{m}\geq 2} S_{n,m} F^*_{\cal O}({m})\,.
\ee
With an abuse of notation, we denoted by $n,m$ some asymptotic state made of $n$ and $m$ particles respectively, so that the sum over $m$ implies an integration over the phase space of the corresponding multi-particle state.
The dilation operator $D\equiv p_i\partial_{p_i}$ can be traded, in massless theories like the ones we consider, with $-\mu\partial_{\mu}$, so that its action on form factors can be deduced from the renormalization group (RG) equations \cite{Caron-Huot:2016cwu}
\be\label{Ddef}
D F_{\cal O}=-\mu\partial_{\mu}F_{\cal O}=(\gamma_{\rm UV}-\gamma_{\rm IR}+\beta_\lambda\partial_\lambda)F_{\cal O}\,.
\ee
Notice that \eq{chw}, as well as \eq{Ddef}, should be interpreted as a matrix equation in the space of all local operators $\cal O$, accounting for possible RG mixing effects. However, since (i) we consider conserved operators, for which $\gamma_{\rm UV}$ is zero, and (ii) IR divergences do not mix form factors (color mixing, while acting non-diagonally in the color space of external states, is diagonal in local-${\cal O}$ space)
, we can safely interpret \eq{chw} to be an equation for the form factor of some fixed $\cal O$, say the energy-momentum tensor $T^{\mu\nu}$.

The version of \eq{chw} we will work with reads as follows
\be\label{master}
e^{i\pi (\gamma_{\rm IR}-\beta_\lambda\partial_\lambda)}F^*_{\cal J}({2})=\sum_{{m}\geq 2} S_{2,m} F^*_{\cal J}({m})\,,
\ee
where $\cal J$ is some conserved current\footnote{In general, that $\cal J$ satisfies a Ward identity is not enough to conclude that $\gamma_{\rm UV}=0$ \cite{Collins:2005nj,Baratella:2021guc}. We will come back to this in due time.}, and the choice $n=2$ corresponds to working with minimal form factors (i.e. those that do not vanish in the free theory limit), which is surely convenient to do. The state `2' will typically be made of a particle-antiparticle pair, so that \eq{master} should be read as an equation for the IR divergences associated to this \emph{state} (at variance with $\gamma_{\rm UV}$, which is a property of the operator $\cal O$)\footnote{In particular, what is here called $\gamma_{\rm coll}$ (see below) is a quantity characterizing the form factor of a particle-antiparticle pair, and its value is \emph{twice} the quantity that is usually called $\gamma_{\rm coll}^{(i)}$ and is associated to the single particle $i$ (and is equal for its antiparticle).}. Once $\gamma_{\rm IR}$ is extracted with \eq{master}, it will enter universally in all processes involving these particles. For notational simplicity, we will omit the suffix ${\cal J}$ from now on.

\vspace{.4cm}

\section{IR renormalization at one loop}\label{1loop}

Expanding \eq{master} to one-loop order, with $S=1+i{\cal M}$, we get
\be\label{oneloop}
\gamma_{\rm IR}^{(1)} \,F_0(2)=\frac{1}{\pi}\,\sum_{2'}\,{\cal M}_{2,2'}^{(0)} \,F_0(2')\,,
\ee
where $F_0(2)\equiv\lim_{\lambda\to 0}F(2)$ is the free theory form factor, which is a real object. By $x^{(\ell)}$ we denote the $\ell$-loop contribution to $x$. The sum over $2'$ denotes a sum over all relevant intermediate two-particle states, integrated over their phase space, while $F_0(2')$ denotes the minimal $\cal J$ form factor over the state $2'$. States with $m\geq 3$ provide a higher-loop effect and drop out of the one-loop equation. More explicitly, \eq{oneloop} can be rewritten as the following two-particle phase space integral against a conventionally normalized amplitude ${\cal A}_{2,2'}$ \cite{Baratella:2020lzz} {(see Appendix \ref{app:norm} for the definition of the $n$-particle phase space in $d$ spacetime dimensions)}
\be\label{gammaIR1loop}
\gamma_{\rm IR}^{(1)} \,F_0(2)=\frac{1}{32\pi^3}\,\sum_{2'}\int_{4\pi}d\Omega_{2'}\,{\cal A}_{2,2'}^{(0)} \,F_0(2')\,.
\ee
This is an equation for $\gamma_{\rm IR}$ at one loop. To extract it, we assume that it has the following structure
\be\label{gammaIRdimreg}
\gamma_{\rm IR}={\gamma_{\rm cusp}}\,\ln \frac{-s}{\mu^2}+\gamma_{\rm coll}\,,
\ee
which is certainly true at one loop in any theory, and conjectured to be valid at all orders and for arbitrary $n$-point amplitudes in QCD \cite{Becher:2009cu,Becher:2014oda} and maximally supersymmetric Yang-Mills \cite{Bern:2005iz}.

With this expression for $\gamma_{\rm IR}$, how do we match the left- and right-hand sides of \eq{oneloop}? Surprisingly enough, we learned from \cite{Caron-Huot:2016cwu} that, in order to make sense of its appearance in \eq{oneloop}, the `cusp' contribution to the infrared anomalous dimension {has to} be identified with a certain phase space integral, specifically
\be\label{gammaIR}
\gamma_{\rm IR}\to \,{\gamma_{\rm cusp}}\int_0^{\pi} d\theta\,s_\theta \,\frac{1}{1-c_\theta}+\gamma_{\rm coll}\,,
\ee
where the polar angle $\theta$ parametrizes, together with the azimuthal angle $\phi$, the phase space of particles $2'$ with respect to some reference frame defined by the 2 system, so that $\theta\to 0$ corresponds to a forward scattering. Notice that the integral in \eq{gammaIR} is divergent.\footnote{\eq{gammaIR} is the appropriate one for distinguishable particles in the cut. For identical particles, we need to substitute $(1-c_\theta)^{-1} \to s^{-2}_\theta$. This is thoroughly discussed in \cite{Baratella:2020dvw}.} By enforcing the validity of \eq{oneloop}, we can then uniquely fix the finite coefficients $\gamma^{(1)}_{\rm cusp}$ and $\gamma^{(1)}_{\rm coll}$. Of course, the non-vanishing of $\gamma^{(1)}_{\rm cusp}$ is possible only when the phase space integral on the right-hand side is also divergent. To make this more concrete, consider scalar QED with a single charged scalar $\phi$. With the methods described in this paper (see the next sections for more details on the various manipulations and a precise definition of the various quantities), and using \eq{gammaIR1loop}, we find the following condition
\begin{align}\label{cuspExample}
    &\left({\gamma^{(1)}_{\rm cusp}}\int_0^{\pi} d\theta\,s_\theta \,\frac{1}{1-c_\theta}+\gamma^{(1)}_{\rm coll}\right){\bf F}^{\mu\nu}_0(1_\phi,2_{\bar{\phi}}) ~~~~~~~~~~~~~~~~~~~~~~~~~~~~~~~~~~~~~~~~~\\
    &~~~~~~~~~~~~~~~~~~~~~~=\frac{1}{32\pi^3}\int_{4\pi}d\Omega_{1'2'}\,{\bf F}^{\mu\nu}_0(1'_\phi,2'_{\bar{\phi}})\,{\cal A}(1'_\phi,2'_{\bar{\phi}}\,;\,1_\phi,2_{\bar{\phi}}) \nonumber \\
    &~~~~~~~~~~~~~~~~~~~~~~={\bf F}^{\mu\nu}_0(1_\phi,2_{\bar{\phi}})\,\,\frac{\,g^2}{16\pi^2}\int_0^{\pi}d\theta \,s_\theta \, d^2_{0,0}(\theta)\left(\,\frac{2}{1-c_\theta}-\frac{c_\theta}{2}-\frac{1}{2} \,\right), \nonumber
\end{align}
which is satisfied if and only if $\gamma_{\rm cusp}^{(1)}=g^2/8\pi^2$ (just look at the coefficient of the divergence in the angular integral for $\theta\to 0$ on both sides of the equation) and $\gamma_{\rm coll}^{(1)}=-3 g^2/8\pi^2$.
This point of view on \eq{oneloop} was not particularly emphasized in previous related works, which mainly interpreted it as an equation for $\gamma_{\rm coll}$, with $\gamma_{\rm cusp}$ taken as an input from
some place else.

\vspace{.5cm}

We may feel unsatisfied about the claimed equivalence of \eq{gammaIR} and \eq{gammaIRdimreg}. A way to formally show their connection is via dimensional regularization, as we are now going to sketch. By promoting the divergent integral multiplying $\gamma_{\rm cusp}$ in \eq{gammaIR} to its regularized version in $4-2\epsilon$ dimensions, we find
\begin{align}\label{8to7}
\int_0^{\pi} \,\frac{d\theta\,s_\theta}{1-c_\theta}~\to ~ n_\epsilon\left(\frac{s}{\mu^2}\right)^{-\epsilon}\int_0^{\pi} \,\frac{\,d\theta\,s_\theta^{1-2\epsilon}\,}{1-c_\theta} =-\frac{1}{\epsilon}+\ln\frac{s}{\mu^2}+\ldots,
\end{align}
where $n_\epsilon$ is a scheme-dependent quantity which tends to 1 for $\epsilon\to 0$. After minimal subtraction, we see that the logarithmic piece of \eq{gammaIRdimreg} is precisely matched.\footnote{This sketch is, of course, just a sketch, and does not address the question of how to treat the finite terms in the ellipses of \eq{8to7}, which naively seem to render the extraction of $\gamma_{\rm coll}$ ambiguous. We refrain from a detailed treatment of cusp anomalies in dimensional regularization in this work, since our main focus is on theories with collinear divergences only.}
In practice, two approaches are possible which of course give equivalent results. Either one adopts \eq{gammaIR} and computes the right-hand-side of \eq{oneloop} in four dimensions (in general with divergences in the phase space integral for $\theta\to 0$) or, alternatively, one can calculate the phase space integral on the right-hand-side of \eq{oneloop} in dimensional regularization, subtract the poles and match the result to $\gamma_{\rm IR}F_0(2)$, with $\gamma_{\rm IR}$ as in \eq{gammaIRdimreg}.

In a sense, the second approach is more satisfactory, as it treats coherently the divergences appearing in phase space integrals and those appearing in loop integrals (which will become relevant when expanding \eq{master} to higher loop orders). However, the important thing in practice is to consistently compare apples to apples and oranges to oranges.

\vspace{.5cm}

Collinear and cusp divergences have a markedly different structure, that allows one to separate them. The first ones are related to a rational contribution to \eq{oneloop}, while the second ones are purely logarithmic. (The possibility of having logarithmic pieces is special to $\gamma_{\rm IR}$.)

There is indeed an elegant way to project out the `cusp' contribution from \eq{oneloop}, which consists in keeping only its rational part. We can express this as
\be\label{gammacoll}
\gamma_{\rm coll}^{(1)} \,F_0(2)=\frac{1}{\pi}\,{\cal R}\,\sum_{2'}\,{\cal M}_{2,2'}^{(0)} \,F_0(2')\,.
\ee
There are several efficient ways to implement the action of $\cal R$, which is defined abstractly as the projector onto the rational part of the function it acts on \cite{Mastrolia:2009dr,Arkani-Hamed:2008owk}. The possibility of separating rational and logarithmic yields can be very useful in practice, since one is often interested in UV divergences of form factors, and not in IR ones. In this case, logarithmic contributions to $\gamma$ can be simply projected out, leaving out, in general, only UV and collinear divergences. Since the two are structurally identical (at one loop, both of them appear as simple poles in dimensional regularization), it is important to be able to separate them. This is where the study of conserved operators, with $\gamma_{\rm UV}=0$, becomes relevant, together with the fact that IR divergences are universal and depend only on long distance physics, i.e. on the particles partaking in the process.

In summary, if one is interested in capturing UV divergences in a massless on-shell scheme, it is crucial to have knowledge of the collinear divergences in order not to mistake them for UV ones. Instead, cusp divergences can be projected out since they are structurally peculiar and cannot be mistaken. We will see a simple example of the subtraction of IR divergences at two loops in Sec.~\ref{sec:quartic}.

\section{Two loops}

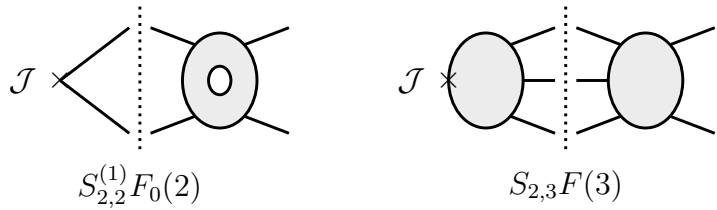
\begin{figure}[t]
\centering
\begin{tikzpicture}[line width=1.1 pt, scale=.7, baseline=(current bounding box.center)]

	\begin{scope}[shift={(7.0,0)}]
	\draw (-1.5,0) to (-.2,1) ;
	\draw (-1.5,0) to (-.2,-1) ;
	\draw (1.5,.5) to  (.2,1) ;
	\draw (1.5,-.5) to  (.2,-1) ;
	\draw (1.5,.5) to  (2.8,1) ;
	\draw (1.5,-.5) to  (2.8,-1) ;
	\node at (-1.5,0) {$\times$} ;
	\node at (-2.2,0) {${\cal J}$} ;
	\draw[dotted] (0,1.4) -- (0,-1.4) ;
	\draw[fill=gray!15] (1.5,0) ellipse (.7 cm and .9 cm) ;
	\draw[fill=white] (1.5,0) ellipse (.21 cm and .27 cm) ;
	\node at (0,-2) {${S}^{(1)}_{2,2}F_{0}(2)$} ;
	\end{scope}

	\begin{scope}[shift={(15.0,0)}]
	\draw (-1.5,0) to (-.2,0) ;
	\draw (-1.5,.5) to  (-.2,1) ;
	\draw (-1.5,-.5) to  (-.2,-1) ;
	\draw (1.5,0) to (.2,0) ;
	\draw (1.5,.5) to  (.2,1) ;
	\draw (1.5,-.5) to  (.2,-1) ;
	\draw (1.5,.5) to  (2.8,1) ;
	\draw (1.5,-.5) to  (2.8,-1) ;
	\draw[fill=gray!15] (-1.5,0) ellipse (.7 cm and .9 cm) ;
	\node at (-2.2,0) {$\times$} ;
	\node at (-2.9,0) {${\cal J}$} ;
	\draw[dotted] (0,1.4) -- (0,-1.4) ;
	\draw[fill=gray!15] (1.5,0) ellipse (.7 cm and .9 cm) ;
	\node at (0,-2) {$S_{2,3}F(3)$} ;
	\end{scope}

 \end{tikzpicture}
 \caption{\emph{Two-loop diagrams entering the right-hand-side of \eq{master2}.}}
 \label{fig:2loop}
 \end{figure}

The aim of this section is to present the general procedure for extracting IR anomalous dimensions up to two loops. The following simple assumption plays a crucial role in simplifying the problem: that the (one-loop) renormalized form factor $F(2)$ can be factorized,\footnote{In general, $C$ can be a matrix acting on the color indices of the external legs of $F_0$. This happens in QCD for example.} i.e.
\be\label{factorizable}
F(2)=C(\lambda)F_0(2)\,,
\ee
where $F_0$ is the free-theory limit, and $\lambda$ is some squared coupling, e.g. $g^2$ in gauge theories (see \cite{Yang:2019vag} for an example in super-Yang-Mills).
Using this, and throwing away contributions from the three-loop order on, we find
\begin{align}
e^{i\pi (\gamma_{\rm IR}-\beta_\lambda \partial_\lambda)}F^*(2)&=\sum_{2'}S_{2,2'} F^*(2')+\sum_{3}S_{2,3} F^*(3)+\ldots \nonumber \\
e^{i\pi (\gamma_{\rm IR}-\beta_\lambda \partial_\lambda)}C^* F_0(2)&=\sum_{2'}C'^* S_{2,2'} F_0(2')+\sum_{3}S_{2,3} F^*(3)+\ldots \nonumber \\
 C^{*-1}\, e^{i\pi (\gamma_{\rm IR}-\beta_{\lambda} \partial_{\lambda})}\, C^*  F_0(2) &=\sum_{2'}\left(\frac{C'}{C}\right)^{\!*} S_{2,2'} F_0(2')+{C}^{*-1}\sum_{3}S_{2,3} F^*(3)+ \ldots \nonumber
\end{align}
Since the three particle cut is already by itself a two-loop effect, and considering that
\be
C=1+a+{\cal O}(\lambda^2)\,,
\ee
(and similarly for $C'$) we get that, up to three-loop effects, the master formula \eq{master} simplifies to
\be\label{master2}
e^{i\pi \left(\gamma_{\rm IR}-\frac{1}{\lambda}\,(a^* +\frac{i\pi}{2}\,\gamma_{\rm IR})\, \beta_\lambda\right)}F_0(2)=\sum_{2'}e^{\left(a'-a\right)^*} S_{2,2'}F_0(2')+\sum_3 S_{2,3}\,F(3)+{\cal O}(\lambda^3)\,,
\ee
where $\beta_\lambda={\cal O}(\lambda^2)$ and $a={\cal O}(\lambda)$, and we dropped the * from $F^*(3)$ since it is a real (tree-level) object in the two-loop approximation. \eq{master2} is the central formula of this work.
Before moving to some examples of its implementation, let us collect the main observations which are general in nature.
\begin{enumerate}
\item The separation in loop orders is pretty clear in \eq{master2}. The zero-loop order is just the identity $F_0=F_0$ (on the right-hand-side, take the 1 from $S_{2,2}=1+i{\cal M}_{2,2}$), while the one-loop order is easily seen to reduce to \eq{oneloop}, as the terms involving $a^*$, $\beta_\lambda$ and $S_{2,3}$ are two-loop by themselves.
\item To go to two loops, \eq{master2} tells us that $F(2)$ should be known up to one loop, including finite terms that cannot be computed with \eq{oneloop}. The relevant information is captured by $a$, which is of the form
\be\label{aeq}
a= a_{\rm fin}^{(1)}-\frac{1}{2}\,\gamma_{\rm coll}^{(1)}\ln\frac{-s-i\varepsilon}{\mu^2}-\frac{1}{4}\,\gamma_{\rm cusp}^{(1)}\ln^{2}\frac{-s-i\varepsilon}{\mu^2}+{\cal O}(\lambda^2)\,.
\ee
\item Besides the full one-loop computation of $F(2)$, the other new ingredients at two loops are (i) the three-particle phase space integral, here schematically written as ${S}_{2,3}F(3)$, with all quantities at tree level, and (ii) the two-particle phase space integrals of $F_0(2')$ with the one-loop order $2\to 2$ amplitudes ${S}^{(1)}_{2,2'}$. This is shown in Fig.~\ref{fig:2loop}.
\item While the zero- and one-loop orders of \eq{master} produce respectively a real and a purely imaginary equation, starting from two loops we can split the equations into a real and an imaginary part, leading to further simplifications on which we will come back when considering the various examples, in particular in Sec.~\ref{SYM}.
\item Depending on the properties specific to the theory under consideration, \eq{master2} can be simplified even further, for example in the maximally supersymmetric Yang-Mills theory, where $\beta_g=0$. In this case, the correction proportional to $\beta$ in the left-hand-side of \eq{master2} is absent, and it is not necessary to compute the one-loop finite terms of $F(2)$.
\end{enumerate}
The two-loop master formula \eq{master2} is powerful, as it allows to compute two-loop RG effects with only one-loop information as an input. This is analogous to the computation of one-loop RG effects from tree-level amplitudes with \eq{oneloop}.

\subsection{$\lambda\varphi^4$ theory}\label{sec:quartic}

As a first example of the method, we now consider the simplest theory of a neutral scalar interacting through a quartic coupling. Using here and in the next section a more explicit notation, we start by listing the relevant objects and making several comments about them. We have, up to one loop\footnote{Here and throughout the paper we adopt dimensional regularization with $\overline{\rm MS}$ subtraction.} (the semicolon separates initial and final states),
\be
{\cal A}(1_\varphi,2_\varphi\,;\,3_\varphi,4_\varphi) = -\lambda+\frac{\lambda^2}{32\pi^2}\left(6 -\ln \frac{-s-i\varepsilon}{\mu^2}- \ln \frac{-t-i\varepsilon}{\mu^2}- \ln \frac{-u-i\varepsilon}{\mu^2}\right)\,,
\ee
where, as usual, $s = (p_1+p_2)^2$, $ t=(p_1-p_3)^2$ and $ u=(p_1-p_4)^2$. The relevant form factor is defined by
\be \label{eq:F2_form_factor}
 _{\rm out}\langle \,1_\varphi, 2_\varphi \,|\,T^{\mu\nu}(x) \,|\, 0\, \rangle=e^{i(p_1+p_2)\cdot x}~ _{\rm out}\langle \,1_\varphi, 2_\varphi \,|\,T^{\mu\nu}(0) \,|\, 0\, \rangle \equiv e^{i (p_1+p_2)\cdot x} \,\,{\bf F}^{\mu\nu}(1_\varphi, 2_\varphi )
\ee
with the further implicit assumption that we project ${\bf F}$ onto its $J=2$ partial wave, for example by taking the outgoing particles to be in some `$d$-wave' state. This extra care is required in order to deal with the fact that $T^{\mu\nu}$ can and in general \emph{does} mix in the UV. To our understanding, however, all mixing effects happen in the $J=0$ sector \cite{Baratella:2021guc} (the wording here refers to the decomposition of operators according to their transformation under $SO(3)$). The reason why we expect no mixing in the $J=2$ sector is the absence of $J=2$ identically conserved operators with which $T^{\mu\nu}$ could mix, unless one turns on gravity. In this case, the energy-momentum tensor can mix with $\partial_\rho \partial_\sigma C^{\mu\rho\nu\sigma}$ (where $C$ is the Weyl tensor), which is purely $J=2$. However, this mixing has to vanish in the limit $M_P \to \infty$, i.e. when gravity is neglected. Notice that, if the renormalization of the form factor were not multiplicative (i.e. if it involved mixing effects), \eq{factorizable} could not have been invoked.

The free-theory form factor, which will play a special role in the following, can be conveniently written as
\be\label{scalarFtau}
{\bf F}_0^{\mu\nu}(1_\varphi, 2_\varphi )=\sqrt{\frac{2}{3}}\,(p_1+p_2)^2\,\tau_{2,0}^{\mu\nu}(\lambda_1,\lambda_2)\,,
\ee
where $\tau_{2,0}$ is the $m=0$ element of a $J=2$ multiplet of tensors, called $\tau_{2,m}$ defined in Appendix~\ref{tautensors}. As usual $\lambda_i$ denotes the spinor helicity variable associated to state $i$, whose momentum is $p_i=\lambda_i\tilde{\lambda}_i$. The normalization of ${\bf F}_0$ is fixed, since the expectation value of $T^{\mu\nu}$ on a state is related to the energy-momentum carried by that state \cite{Caron-Huot:2016cwu}.

With these ingredients, let us now compute the right-hand side of \eq{master2} in the $\varphi^4$ theory. We first observe that $S_{2,3}=0$, so that the last term of \eq{master2} vanishes. Second, to greatly simplify the two-particle phase space integral, we note that the free-theory form factor can be written, in angular coordinates, as (see \eq{tauangular})
\be\label{ffangular}
{\bf F}_0^{\mu\nu}(1'_\varphi, 2'_\varphi )=s\,\sqrt{\frac{2}{3}}\sum_{m=-2}^2 \,\tau_{2,m}^{\mu\nu}(\lambda_1,\lambda_2)\,d^2_{m,0}(\theta)\,e^{im\phi}={\bf F}_0^{\mu\nu}(1_\varphi, 2_\varphi )\,d^2_{0,0}(\theta)+\ldots \,,
\ee
where the primed variables are given by
\begin{align}\label{lambdaprime}
    \lambda'_1&= c_{\frac{\theta}{2}}\lambda_1-s_{\frac{\theta}{2}} e^{-i\phi}\lambda_2 \nonumber \\
    \lambda'_2&= s_{\frac{\theta}{2}}e^{i\phi}\lambda_1+c_{\frac{\theta}{2}}\lambda_2
\end{align}
and the ellipses in \eq{ffangular} stand for terms that have a non-trivial dependence on the azimuthal angle $\phi$.

Up to three-loop corrections, we then have (the factor 1/2 in the phase space integral is for identical internal particles)
\begin{align}\label{scalar}
e^{i\pi \left(\gamma_{\rm IR}-\frac{1}{\lambda}\,(a^* +\frac{i\pi}{2}\,\gamma_{\rm IR})\, \beta_\lambda\right)}{\bf F}_0^{\mu\nu}(1_\varphi, 2_\varphi )&={\bf F}_0^{\mu\nu}(1_\varphi, 2_\varphi )+\frac{i}{64\pi^2}\int_{4\pi}s_\theta d\theta d\phi ~{\bf F}_0^{\mu\nu}(1'_\varphi, 2'_\varphi )\,
{\cal A}(1'_\varphi,2'_\varphi;1_\varphi,2_\varphi) \nonumber \\
&={\bf F}_0^{\mu\nu}(1_\varphi, 2_\varphi )\left(1+\frac{i}{32\pi}\int_0^\pi s_\theta d\theta~d^2_{0,0}(\theta)\,{\cal A}(\theta) \right) \nonumber \\
&={\bf F}_0^{\mu\nu}(1_\varphi, 2_\varphi )\left(1+\frac{i}{32\pi}\int_0^\pi s_\theta d\theta~d^2_{0,0}(\theta)\left[\Delta-\frac{\lambda^2}{16\pi^2}\ln s_\theta\right] \right) \nonumber \\
&={\bf F}_0^{\mu\nu}(1_\varphi, 2_\varphi )\left(1+\frac{i\pi\lambda^2}{6(16\pi^2)^2}\right).
\end{align}
In the second equality we expressed ${\cal A}$ with angular variables thanks to \eq{lambdaprime}, while in the third equality we introduced
\be
\Delta \equiv -\lambda+\frac{\lambda^2}{32\pi^2}\left(6-3\ln\frac{s}{\mu^2}+i\pi+2\ln 2\right)\,.
\ee 
Since $\Delta$ has no angular dependence, it is annihilated by the $d^2$ phase space integral. Notice that this includes the one-loop contribution to \eq{master2} (coming from the tree-level contribution to ${\cal A}$), the two-loop contribution coming from the one-loop rational
term of $\cal A$, and finally the two-loop contribution
coming from the $\ln\mu$ term of $\cal A$.
Before extracting the result, let us make some more remark about \eq{scalar}.
\begin{enumerate}
\item Perhaps the most important point is that there is a certain degree of ``bootstrapping'' in \eq{scalar}. First of all, in the fact that both sides of the equation are proportional to the same (free-theory) tensor ${\bf F}_0^{\mu\nu}$. In practice, this is guaranteed by the $d\phi$ integration, which projects out all the unwanted structures (see \eq{ffangular}).
\item More interestingly, we know that the physics that feeds $a^*$ includes contributions to $\gamma_{\rm IR}$ at one loop, which are non-local (see \eq{aeq}). If these were non-zero (and since $\beta^{(1)}_\lambda$ does not vanish in this theory), we would need similar non-local two-loop contributions on the right-hand side. But we find just a number. Consistently, we see that $a$ is indeed finite, since the one-loop contribution in the right-hand side is zero (it is all in $\Delta$). In this respect, $\varphi^4$ is special.
 \item Since $a^*$ contains a finite, scheme dependent term, the two-loop anomalous dimension will in general be scheme dependent as well. The scheme dependence is precisely dictated by \eq{scalar}, and in general by \eq{master2}. However, as we said, by expanding \eq{scalar} we get, at one loop, $\gamma_{\rm IR}^{(1)}=0$. Since there is no need for a renormalization at one loop (there are no UV nor IR divergences) $a\equiv a_{\rm fin}^{(1)}$ is physical. In order not to spoil the {universal properties of the energy-momentum tensor}, it must vanish.\footnote{We checked that $a=0$ at one-loop order by an explicit computation.}
\end{enumerate}
We therefore get at two loops
\be
\gamma_{\rm IR}^{(2)}=\frac{\lambda^2}{6(16\pi^2)^2}\,.
\ee
Because we obtain a pure number, we rediscover that only collinear divergences are present up to two loops in $\lambda \varphi^4$.

\subsubsection{Operator self-renormalization}

In order to demonstrate how operator anomalous dimensions are extracted in the presence of IR divergences, let us consider a simple variation of the previous example. We take a charged scalar $\phi$ and study loop effects governed by the quartic coupling ${\lambda'}|\phi|^4$. The only notable difference with respect to the neutral scalar is that now the $2\to 2$ amplitude is not fully crossing symmetric
\be
{\cal A}(1_\phi,2_{\Bar{\phi}}\,;\,3_\phi,4_{\Bar{\phi}})=-\lambda'+\frac{\lambda'^2}{16\pi^2}\left(5-\ln\frac{-s}{\mu^2}-\ln\frac{-t}{\mu^2}-\frac{1}{2}\ln\frac{-u}{\mu^2}\right)+{\cal O}(\lambda'^3)\,,
\ee
and the IR anomalous dimension is equal to
\be\label{gammaIRcharged}
\gamma_{\rm coll}|_\phi = -\frac{ \lambda'^2}{(16\pi^2)^2}\int_0^\pi s_\theta d\theta~d^2_{0,0}(\theta) \left(\ln(1-c_\theta)+\frac{1}{2}\ln(1+c_\theta)\right)=\frac{ \lambda'^2}{2(16\pi^2)^2}\,.
\ee
Let us now consider the higher dimensional operator ${\cal O}_{\psi\phi}$ involving $\phi$ and a charged fermion $\psi$, whose free-theory form factor is given by
\be\label{ffO}
{\bf F}_{{\cal O}_{\psi\phi}}(1_{\psi_+},2_{\psi_-},3_\phi,4_{\Bar{\phi}})=\la 1 |(p_3-p_4)| 2]\,,
\ee
and study its self-renormalization as induced by $\lambda'$ effects. A one-loop computation reveals that no correction to ${\bf F}_{{\cal O}_{\psi\phi}}$ arises at this order, so that \eq{chw} can be simplified to
\begin{align}
\left(\gamma^{(2)}_{{\cal O}_{\psi\phi}}-\gamma^{(2)}_{\rm IR}\right)&{\bf F}_{{\cal O}_{\psi\phi}}(1_{\psi_+},2_{\psi_-},3_\phi,4_{\Bar{\phi}})=\frac{-1}{32\pi^3}\int_{4\pi}d\Omega_{3'4'} ~{\bf F}_{{\cal O}_{\psi\phi}}(1_{\psi_+},2_{\psi_-},3'_\phi,4'_{\Bar{\phi}})\,{\cal A}^{(1)}(3'_\phi,4'_{\Bar{\phi}}\,;\,3_\phi,4_{\Bar{\phi}}) \nonumber \\
&={\bf F}_{{\cal O}_{\psi\phi}}(1_{\psi_+},2_{\psi_-},3_\phi,4_{\Bar{\phi}})\,\frac{\lambda'^2}{(16\pi^2)^2}\int_0^\pi s_\theta d\theta~d^1_{0,0}(\theta) \left(\ln(1-c_\theta)+\frac{1}{2}\ln(1+c_\theta)\right)\nonumber \\
&={\bf F}_{{\cal O}_{\psi\phi}}(1_{\psi_+},2_{\psi_-},3_\phi,4_{\Bar{\phi}})\,\frac{-\lambda'^2}{{2(16\pi^2)^2}}\,.
\end{align}
Plugging in the value of $\gamma_{\rm IR}$ that is appropriate for this process, which is precisely given by \eq{gammaIRcharged}, we get that $\gamma^{(2)}_{\cal O}=0$. At first sight, it might seem that there is an accidental cancellation due to a numerical coincidence in the values of $\int  d^J(\theta) {\cal A}(\theta)$ for $J=1,2$ (notice that no other pair of values for $J$ gives the same integral). However, there is a deeper reason behind this null result. The operator which gives rise to \eq{ffO} is ${\cal O}_{\psi\phi}={\psi}^\dagger \sigma^\mu\psi (\Bar{\phi}{{\partial}}_\mu\phi-{\phi}{{\partial}}_\mu\Bar{\phi})$, i.e. it is a product of conserved currents. Since fermions are inert under $\lambda'$ interactions, they act like spectators in the renormalization of ${\cal O}_{\psi\phi}$. This means, in other words, that we just studied the self-renormalization of $j_\mu=\Bar{\phi}{{\partial}}_\mu\phi-{\phi}{{\partial}}_\mu\Bar{\phi}$, even if disguised. Our null result appears then as a welcome cross-check of the consistency of the method, because the extraction of $\gamma_{\rm IR}$ in no way should depend on the conserved current that is being considered in \eq{master}. As a last comment, we notice that, while it is not always possible to define a conserved charge and the associated $j^\mu$, the energy momentum tensor $T^{\mu\nu}$ is always available (like for the neutral $\varphi$).

\subsection{Yukawa theory} \label{sec:yukawa}

In this section we would like to consider an example which, from the point of view of \eq{master2}, is slightly more complicated, in that both $\beta^{(1)}$ and $a^{(1)}$ are different from zero, with $a^{(1)}$ being non-local due to the presence of collinear divergences at one loop: the Yukawa theory of a neutral scalar $\varphi$ and a {Dirac} fermion $(\psi_\pm,\psitil_\pm )$.\footnote{$\psi$ and $\psitil$ represent two distinct particle-antiparticle pairs, while their index denotes the sign of helicity. One can assign to $\psi_+$ some charge $Q$ and to $\psitil_+$ its negative.} The goal is to compute the IR anomalous dimension in the scalar sector at two loops using \eq{master2}. This means that state `2' will be made of a pair of scalars. All the results presented in this section, which are based on \eq{master2}, have been cross checked with traditional methods -- see Appendix~\ref{appC}. The calculation for the form factor with external fermions proceeds in the same way and is presented in Appendix \ref{app:fermion_sector}. Additionally, we show a non-trivial cross check for fermion sector with the help of an auxiliary gauge field.

We start by listing all the ingredients that are necessary for the evaluation of the left- and right-hand sides of the master equation. First of all, we need the complete one-loop form factors with two external scalars and two external fermions, given respectively by
\begin{align}
  {\bf F}^{\mu\nu}(1_\varphi, 2_\varphi )&={\bf F}_0^{\mu\nu}(1_\varphi, 2_\varphi )\,\left[1+\frac{y^2}{16\pi^2}\left(6-2\ln\frac{-s-i\varepsilon}{\mu^2}\right)\right], \label{formfactors} \\
 {\bf F}^{\mu\nu}(1_{\psi_-}, 2_{\psi_+})&={\bf F}_0^{\mu\nu}(1_{\psi_-}, 2_{\psi_+})\,\left[1+\frac{y^2}{16\pi^2}\left(\frac{3}{2}-\frac{1}{2}\ln\frac{-s-i\varepsilon}{\mu^2}\right) \right], \label{formfactorf}
\end{align}
and similarly for $\psitil$. In particular we find
\begin{align}
    a^{(1)}_\varphi&=\frac{y^2}{16\pi^2}\left(6-2\,\ln\frac{-s-i\varepsilon}{\mu^2}\right) , \\
    a^{(1)}_\psi=a^{(1)}_{\psitil} &=\frac{y^2}{16\pi^2}\left(\frac{3}{2}-\frac{1}{2}\,\ln\frac{-s-i\varepsilon}{\mu^2}\right),
\end{align}
from which, by comparison with \eq{aeq}, we can extract the one-loop collinear anomalous dimensions in the scalar and fermion sectors, given respectively by $\gamma^{(1)}_{\rm coll}|_\varphi=4\,(y^2/16\pi^2)$ and $\gamma^{(1)}_{\rm coll}|_\psi=y^2/16\pi^2$. We also see that $\gamma_{\rm cusp}^{(1)}=0$ in both sectors. Notice that in \eq{formfactors} and \eq{formfactorf} we kept the Feynman $i\varepsilon$: when using \eq{master}, one has to change its sign in order to account for the complex conjugation of the form factors dictated by the equation. We will sometimes omit the $i\varepsilon$ in writing down amplitudes. To restore it, one should simply substitute $s$ with $s+i\varepsilon$, and similarly with $t$ and $u$.

We also notice that the free-theory fermionic form factor can be expressed as
\be
{\bf F}_0^{\mu\nu}(1_{\psi_-}, 2_{\psi_+})=(p_1+p_2)^2\,\tau_{2,-1}^{\mu\nu}(\lambda_1,\lambda_2)\,,
\ee
with $\tau_{2,-1}$ given in Appendix~\ref{tautensors}. The relevant amplitudes, complete up to the one-loop order, are given by
\begin{align}
&{\cal A}(1_\varphi, 2_\varphi \,;\,3_\varphi ,4_\varphi )=\frac{-y^4}{8\pi^2} \left[3\pi^2 +\ln^2 \frac{s}{t}+ \ln^2 \frac{t}{u}+\ln^2 \frac{u}{s} +4\left(6-\ln \frac{-s}{\mu^2}- \ln \frac{-t}{\mu^2}- \ln \frac{-u}{\mu^2} \right) \right], \label{scalar1loop}\\
&{\cal A}(1_{\psi_-},2_{\psi_+} \,;\, 3_\varphi, 4_\varphi)
=\frac{\langle 13\rangle}{\langle 23 \rangle} \left[y^2+\frac{y^4}{32\pi^2}\left(\pi^2 +\ln^2 \frac{s}{t}\right)-\frac{5y^4}{32\pi^2} \left( 2-\ln \frac{-u}{\mu^2}\right)\right]+3\leftrightarrow 4\, . \label{fermionscalar}
\end{align}
Notice that \eq{scalar1loop} has no tree-level contribution, which leads to inconsistencies (related to the fact that we have not included a $\lambda\varphi^4$ interaction, which is renormalizable and allowed by symmetries). However, in light of the discussion after \eq{scalar}, we see that such a tree-level term would not play any role in our computation, being just a constant with no angular dependence.

By imposing the RG equations to \eq{fermionscalar}, we can deduce the beta function of the coupling $y$ at one loop. For amplitudes that are defined by minimal subtraction in the UV and in the IR, like the ones we consider here, the appropriate RG equation reads
\be
\left(\mu\partial_\mu-\gamma_{\rm IR}+\beta_y\partial_y\right){\cal A}=0\,,
\ee
with $\gamma_{\rm IR}=\gamma^{(1)}_{\rm coll}|_\varphi+\gamma^{(1)}_{\rm coll}|_\psi$ when applying the RG equation to \eq{fermionscalar}. We find that
\be
\beta_y=\frac{5\, y^3}{16\pi^2},
\ee
to one-loop accuracy.

In order to compute the last contribution to \eq{master2}, which is purely two-loop, we need the following $3\to 2$ amplitude, compactly written as
\be
{\cal A}(1_{\psi_{+}},2_{\psitil_+},3_\varphi\,;\, 4_\varphi, 5_\varphi)={-}\, y^3 \,\sum_{\sigma} \,\frac{\la {\sigma}(3){\sigma}(4) \ra}{\la 1{\,\sigma}(3)\ra \la 2\,{\sigma}(4) \ra}\,,
\ee
where ${\sigma}$ runs over the permutations acting on the scalars $3,4,5$. This amplitude has to be convoluted with the form factor defined by
\be \label{tmunu3}
    _{\rm out}  \la\,  1_{\psi_+},2_{\psitil_+},3_\varphi \,|\,T^{\mu\nu}(x)\,|\,0\,\ra = e^{i(p_1+p_2+p_3)\cdot x} \,\,{\bf F}^{\mu\nu}(1_{\psi_+}, 2_{\psitil_+},3_\varphi),
\ee
where ${\bf F}^{ab\dot{a}\dot{b}}\equiv\sigma_{\mu}^{a\dot{a}}\sigma_{\nu}^{b\dot{b}}\,{\bf F}^{\mu\nu}$ ($\sigma_\mu=(\mathds{1},\sigma_i)$ as usual) is given, with indices $ab\dot{a}\dot{b}$ omitted, by
\begin{align}
{\bf F}&=\, y\left[\,\frac{2}{3}\la 12 \ra \, \epsilon \otimes \tilde{\epsilon} -\frac{1}{3[21]} \left( \lambda_1\lambda_1 \tilde{\lambda}_1\tilde{\lambda}_1+\lambda_1\lambda_2 \tilde{\lambda}_1\tilde{\lambda}_2+\lambda_2\lambda_1 \tilde{\lambda}_2\tilde{\lambda}_1+\lambda_2\lambda_2 \tilde{\lambda}_2\tilde{\lambda}_2+\lambda_3\lambda_3 \tilde{\lambda}_3\tilde{\lambda}_3 \right.\right. \nonumber \\
&\left.  -2\, \lambda_1\lambda_3 \tilde{\lambda}_1\tilde{\lambda}_3-2\,\lambda_2\lambda_3 \tilde{\lambda}_2\tilde{\lambda}_3-2\,\lambda_3\lambda_1 \tilde{\lambda}_3\tilde{\lambda}_1-2\,\lambda_3\lambda_2 \tilde{\lambda}_3\tilde{\lambda}_2+(s_{13}+s_{23})\,\epsilon \otimes \tilde{\epsilon}\,  \right)\nonumber \\
&+\frac{1}{2[31]}\left(\lambda_2\lambda_2 \tilde{\lambda}_2\tilde{\lambda}_3-\lambda_1\lambda_2 \tilde{\lambda}_1\tilde{\lambda}_3-\lambda_3\lambda_2 \tilde{\lambda}_3\tilde{\lambda}_3+\lambda_2\lambda_2 \tilde{\lambda}_3\tilde{\lambda}_2-\lambda_2\lambda_1 \tilde{\lambda}_3\tilde{\lambda}_1-\lambda_2\lambda_3 \tilde{\lambda}_3\tilde{\lambda}_3\right)\nonumber \\
&\left. -\frac{1}{2[32]}\left(\lambda_1\lambda_1 \tilde{\lambda}_1\tilde{\lambda}_3-\lambda_2\lambda_1 \tilde{\lambda}_2\tilde{\lambda}_3-\lambda_3\lambda_1 \tilde{\lambda}_3\tilde{\lambda}_3+\lambda_1\lambda_1 \tilde{\lambda}_3\tilde{\lambda}_1-\lambda_1\lambda_2 \tilde{\lambda}_3\tilde{\lambda}_2-\lambda_1\lambda_3 \tilde{\lambda}_3\tilde{\lambda}_3\right)\right]\,,
\end{align}
with $\epsilon \otimes \tilde{\epsilon}\equiv \epsilon^{ab}\epsilon^{\dot{a}\dot{b}}$. The tensor ${\bf F}^{ab\dot{a}\dot{b}}$ has several properties: (i) it is symmetric in both $ab$ and $\dot{a}\dot{b}$, i.e. it satisfies $\epsilon_{ab}\,{\bf F}^{ab\dot{a}\dot{b}}=\epsilon_{\dot{a}\dot{b}}\,{\bf F}^{ab\dot{a}\dot{b}}=0$ ; (ii) it is transverse, in that it satisfies $(p_1+p_2+p_3)_{a\dot{a}}\,{\bf F}^{ab\dot{a}\dot{b}}=0$ ; (iii) it has the standard properties that on-shell amplitudes satisfy, specifically a well defined little group scaling and a well defined pole structure, with residues fixed by consistent factorization. 

Thanks to properties (i) and (ii), ${\bf F}$ admits an expansion over the multiplet $\tau_{2,m}$ defined in Appendix~\ref{tautensors}, that is
\be\label{Fontau}
{\bf F}^{ab\dot{a}\dot{b}}\!\left( 1'_{\psi_+},2'_{\psitil_+},3'_\varphi\right)=\sum_{m=-2}^2 \tau_{2,m}^{ab\dot{a}\dot{b}}(\lambda_1,\lambda_2)\, {\bm f}_m(\vartheta_i)\,,
\ee
where $\{\vartheta_i\}$ is some set of 5 parameters that allows to express $\lambda'_1,\lambda'_2$ and $\lambda'_3$ in terms of $\lambda_1$ and $\lambda_2$, as discussed for example in \cite{Caron-Huot:2016cwu,EliasMiro:2020tdv}. One can think of the functions ${\bm f}_m$ as some set of `reduced form factors', which are enough to recontruct the full form factor but are simpler objects, i.e. just functions of the relevant relativistic invariants instead of a multi-index tensor. Thanks to the orthogonality properties of the $\tau_{2,m}$ tensors, we also have the inverse formula
\be
{\bm f}_m(\vartheta_i)=(-1)^m \,\tau_{2,-m}^{ab\dot{a}\dot{b}}(\lambda_1,\lambda_2)\,{\bf F}_{ab\dot{a}\dot{b}}\!\left( 1'_{\psi_+},2'_{\psitil_+},3'_\varphi\right) .
\ee
We will come back to the usefulness of these comments when computing the three-particle phase space integral in \ref{sec:3pPSI}, where we will see that only ${\bm f}_0$ matters for the computation.

Notice that here and in the following sections we use ${\bf F}^{\mu\nu}$ and ${\bf F}^{ab\dot{a}\dot{b}}$ interchangeably. The choice is based on which notation gives a more compact or elegant expression, and one can go back and forth from one form to the other by the use of the $\sigma_\mu^{a\dot{a}}$ matrices, paying attention to factors of 2 coming from $\sigma_\mu^{a\dot{a}}\,{\bar{\sigma}}^\nu_{\dot{a}a}=2\,\delta_\mu^\nu$.

\vspace{.5cm}

Before moving to the computation of the relevant cuts, we present the version of \eq{master2} that is appropriate for the extraction of $\gamma_{\rm IR}^{(2)}|_\varphi$
\begin{align}\label{masterYukawa}
    e^{i\pi \left(\gamma_{\rm IR}|_\varphi-\frac{2}{y}\,(a_\varphi^* +\frac{i\pi}{2}\,\gamma_{\rm IR}|_\varphi)\, \beta_y\right)}\,&{\bf F}_0^{\mu\nu}\!\left(1_\varphi, 2_\varphi \right)=\,{\bf F}_0^{\mu\nu}\!\!\left(1'_\varphi, 2'_\varphi \right)S\!\left(1'_\varphi, 2'_\varphi \,;\, 1_\varphi, 2_\varphi \right) \\
    &+\left(e^{\left(a_{\psi}-\,a_{\varphi}\right)^{\!*}}{\bf F}_0^{\mu\nu}\!\!\left(1'_{\psi_-}, 2'_{\psi_+} \right)S\!\left(1'_{\psi_-}, 2'_{\psi_+} \,;\, 1_\varphi, 2_\varphi \right) + \left(\psi \to \psitil\right)\right) \nonumber \\
    &+\sum_{\pm}  \, {\bf F}^{\mu\nu}\!\!\left(1'_{\psi_\pm}, 2'_{\psitil_\pm},3'_\varphi\right)\, S\!\left(1'_{\psi_\pm}, 2'_{\psitil_\pm},3'_\varphi\,;\,1_\varphi, 2_\varphi\right)+\,{\cal O}(y^6)\,.\nonumber
\end{align}
The repeated primed kinematic variables are to be integrated over as usual.

\subsubsection{Two-particle phase space integrals}

The two-particle phase space integrals in \eq{masterYukawa} are staightforward to compute in the Yukawa theory, especially thanks to the absence of `small angle divergences', which would require an infinite subtraction like in \eq{cuspExample}. We will compute the integrals by taking advantage of the angular parametrization that we developed before. We find
\begin{align}
{\bf F}_0^{\mu\nu}\!\!\left(1'_\varphi, 2'_\varphi \right)S\!\left(1'_\varphi, 2'_\varphi \,;\, 1_\varphi, 2_\varphi \right)&={\bf F}_0^{\mu\nu}(1_\varphi, 2_\varphi )+\frac{i}{64\pi^2}\int_{4\pi}s_\theta d\theta d\phi ~{\bf F}_0^{\mu\nu}(1'_\varphi, 2'_\varphi )\,
{\cal A}(1'_\varphi,2'_\varphi;1_\varphi,2_\varphi) \nonumber \\
&={\bf F}_0^{\mu\nu}(1_\varphi, 2_\varphi )\left(1+\frac{i}{32\pi}\int_0^\pi s_\theta d\theta~d^2_{0,0}(\theta){\cal A}_{2\varphi\to 2\varphi}(\theta) \right) \nonumber \\
&={\bf F}_0^{\mu\nu}(1_\varphi, 2_\varphi )\left[1+i\pi \frac{y^4}{(16\pi^2)^2} \left( \!-\frac{94}{9}+ i\pi\, \frac{4}{3}\,\right) \right],
\end{align}
where the four-scalar amplitude is given, in angular coordinates and up to one-loop order, by
\begin{align}\label{phi2phiangle}
{\cal A}_{2\varphi\to 2\varphi}(\theta)&=-\frac{y^4}{8\pi^2}\left[ 24 - 12 \ln\frac{s}{\mu^2}-4\ln\frac{1-c_\theta^2}{4}+ 4\pi i\right. \nonumber \\
&\left. +\ln^2\frac{1-c_\theta}{2}+\ln^2\frac{1+c_\theta}{2}+ 2\pi i\ln\frac{1-c_\theta^2}{4}+\pi^2+\ln^2\frac{1+c_\theta}{1-c_\theta} \right]\,.
\end{align}
The first line accounts for the rational term and the bubble, while the second line for the ``IR-regulated'' box contribution \cite{Baratella:2020lzz}. \eq{phi2phiangle} is obtained from \eq{scalar1loop} by first substituting $1,2$ with $1',2'$ and $3,4$ with $1,2$ and then using \eq{lambdaprime}.

The contribution coming from the exchange of two fermions can be treated in a similar way, knowing that (thanks to \eq{tauangular})
\begin{align}\label{ffpsiangular}
{\bf F}_0^{\mu\nu}(1'_{\psi_-}, 2'_{\psi_+} )&=s\,\tau_{2,-1}^{\mu\nu}(\lambda'_1,\lambda'_2)=s\sum_{m=-2}^2 \,\tau_{2,m}^{\mu\nu}(\lambda_1,\lambda_2)\,d^2_{m,-1}(\theta)\,e^{i(m+1)\phi} \nonumber \\
&={\bf F}_0^{\mu\nu}(1_\varphi, 2_\varphi )\,e^{i\phi}\,\sqrt{\frac{3}{2}}\,d^2_{0,-1}(\theta)+\ldots\,,
\end{align}
the ellipses standing for terms with a different $\phi$ dependence, which are going to be annihilated by the $d\phi$ integration. The second term of \eq{masterYukawa} therefore yields
\begin{align}
{\bf F}_0^{\mu\nu}\!\!\left(1'_{\psi_-}, 2'_{\psi_+} \right)S\!\left(1'_{\psi_-}, 2'_{\psi_+} \,;\, 1_\varphi, 2_\varphi \right)&=\frac{i}{32\pi^2}\int_{4\pi}s_\theta d\theta d\phi ~{\bf F}_0^{\mu\nu}\!\!\left(1'_{\psi_-}, 2'_{\psi_+} \right){\cal A}\!\left(1'_{\psi_-}, 2'_{\psi_+} \,;\, 1_\varphi, 2_\varphi \right)  \\
&={\bf F}_0^{\mu\nu}(1_\varphi, 2_\varphi )\frac{i}{16\pi}\sqrt{\frac{3}{2}}\int_0^\pi s_\theta d\theta~d^2_{0,-1}(\theta){\cal A}_{2\psi\to 2\varphi}(\theta,0) \nonumber \\
&={\bf F}_0^{\mu\nu}(1_\varphi, 2_\varphi )\,\frac{i\pi\, y^2}{16\pi^2}\left[ 2+\frac{ y^2}{16\pi^2}\left(5\ln\frac{s}{\mu^2}-\frac{223}{9}+\frac{i\pi}{3}\right)\right], \nonumber
\end{align}
where the relevant $2\to 2$ amplitude is given in angular coordinates by
\begin{align}
    {\cal A}_{2 \psi \to 2\varphi}(\theta,\phi)=&-y^2\, t_{\frac{\theta}{2}}
    \, e^{-i\phi}\left[ 1 - \frac{y^2}{32\pi^2} \right.  \left. \left( 10 -5 \ln \frac{s}{\mu^2} -5\ln \frac{1+c_\theta}{2} \right. \right. 
     \\
     & \left. \left.-\ln^2 \frac{1-c_\theta}{2} -2\pi i \ln \frac{1-c_\theta}{2}  \right) \right]-\left(\theta\to \pi-\theta\right)\,. \nonumber
\end{align}
An identical contribution comes from the term that has a pair of $\psitil$ instead of $\psi$ as internal states, that is ${\bf F}_0^{\mu\nu}(1'_{\psitil_-}, 2'_{\psitil_+} )\,S(1'_{\psitil_-}, 2'_{\psitil_+} \,;\, 1_\varphi, 2_\varphi )$.

\subsubsection{Three-particle phase space integral}\label{sec:3pPSI}

The remaining contributions to \eq{masterYukawa} are of the following form
\begin{align}
    {\bf F}^{\mu\nu}\!\left(1', 2',3'\right)\, S\!\left(1', 2',3'\,;\,1, 2\right)&=i \int d\Phi_3^{(4)}\,{\bf F}^{\mu\nu}\!\left(1', 2',3'\right) {\cal A}\!\left(1', 2',3'\,;\,1, 2\right) \nonumber \\
    &=\frac{i\pi \,s}{(16\pi^2)^2}\int d^5\vartheta \,\,{\bf F}^{\mu\nu}\!\left(1', 2',3'\right)  {\cal A}\!\left(1', 2',3'\,;\,1, 2\right) \nonumber \\
    &=\frac{i\pi \,s}{(16\pi^2)^2}\sum_{m}\tau_{2,m}^{\mu\nu}(1,2)\int d^5\vartheta \,\, {\bm f}_m(\vartheta)\, {\cal A}_{3\to 2}(\vartheta) \nonumber \\
    &={\bf F}_0^{\mu\nu}(1,2)\,\frac{i\pi}{(16\pi^2)^2}\sqrt{\frac{3}{2}}\int d^5\vartheta \,\, {\bm f}_0(\vartheta)\, {\cal A}_{3\to 2}(\vartheta)\,.~~~~
\end{align}
In the first and second equality we respectively used the definition of the convolution over the primed variables and formally rewritten the phase space integral as a standard integral over 5 parameters with unit measure, i.e. with $\int d^5\vartheta=1$ (examples will be given soon). In the third equality we have expanded ${\bf F}$ over the basis $\tau_{2,m}$ following \eq{Fontau} while, in the last equality, we have used that $\int d^5\vartheta {\bm f}_m(\vartheta){\cal A}(\vartheta)=0$ unless $m=0$. This is a consequence of angular momentum conservation and the helicity structure of ${\cal A}$, and is crucial to ensure that the final result is proportional to the free-theory form factor ${\bf F}_0$.

The validity of the above manipulations can be proven in our specific example using the following parametrization \cite{Caron-Huot:2016cwu,EliasMiro:2020tdv}
\begin{align}\label{param2}
\lambda'_1&=c_{\theta_2} \lambda_1 - e^{-i\phi} c_{\theta_1}s_{\theta_2} \lambda_2 \,, \nonumber \\
\lambda'_2&=s_{\theta_2}c_{\theta_3} \lambda_1 + e^{-i\phi} \left(c_{\theta_1}c_{\theta_2}c_{\theta_3} -e^{-i\rho}s_{\theta_1}s_{\theta_3}\right)\lambda_2\,,  \nonumber \\
\lambda'_3&=s_{\theta_2}s_{\theta_3} \lambda_1 + e^{-i\phi} \left(c_{\theta_1}c_{\theta_2}s_{\theta_3} +e^{-i\rho}s_{\theta_1}c_{\theta_3}\right)\lambda_2 \,,
\end{align}
with measure
\be
\int d^5\vartheta\equiv \frac{4}{\pi^2}\,\int_0^{\pi/2} s_{\theta_1}c_{\theta_1}d\theta_1 \int_0^{\pi/2} s^3_{\theta_2}c_{\theta_2}d\theta_2 \int_0^{\pi/2} s_{\theta_3}c_{\theta_3}d\theta_3 \int_0^{2\pi} {d\rho}\int_0^{2\pi} {d\phi}=1\,.
\ee
In terms of these coordinates, it can be shown that
\be
{\cal A}\!\left(1'_{\psi_{+}},2'_{\psitil_+},3'_\varphi\,;\, 1_\varphi, 2_\varphi\right)=\frac{y^3 \,e^{i\phi}}{\la 12 \ra} \,\, a\!\left(\theta_1,\theta_2,\theta_3,e^{i\rho}\right)\,,
\ee
while
\be
{\bf F}^{\mu\nu}\!\left(1'_{\psi_{+}},2'_{\psitil_+},3'_\varphi\right)=y\,\la 12\ra\,\sum_m \tau^{\mu\nu}_{2,m}(\lambda_1,\lambda_2) \, e^{i(m-1)\phi} \, t_m\!\left(\theta_1,\theta_2,\theta_3,e^{i\rho}\right)\,,
\ee
with $a$ and $t_m$ some dimensionless functions of the specified parameters. This shows that, after $d\phi$ integration, only $\tau_{2,0}$ survives as anticipated.

To see how the computation of a three-particle phase space integral works in practice, we consider an explicit example, and compute the following diagram
\begin{align}\label{3cutexample}
    &\begin{tikzpicture}[line width=.7 pt, scale=0.6, baseline=(current bounding box.center)
    ]
	\draw (-1.2,0) to (-.1,1) ;
	\draw (-1.2,0) to (-.1,-1) ;
	\draw[dashed]  (-.65,.5) to  (-.1,0) ;
	\draw[dotted] (0,1.4) -- (0,-1.4) ;
    \draw (.1,1) -- (1,.5) -- (1,-.5) -- (.1,-1) ;
    \draw[dashed] (.1,0) to (1,0) ;
    \draw[dashed] (1,.5) to (2,1) ;
    \draw[dashed] (1,-.5) to (2,-1) ;
	\node at (-1.95,.2) {\footnotesize $T^{ab\dot{a}\dot{b}}$} ;
    \node at (2.2,-.8) {\tiny $1$} ;
    \node at (2.2,.8) {\tiny $2$} ;
    \node at (.4,-1.2) {\tiny $1'$} ;
    \node at (.4,.35) {\tiny $3'$} ;
    \node at (.4,1.3) {\tiny $2'$} ;
 \end{tikzpicture}~~=\frac{i\pi \,s}{(16\pi^2)^2}\int d^5\vartheta\,\frac{-y}{2[3'2']}\left(\lambda_{1'}^a\lambda_{1'}^b \tilde{\lambda}_{1'}^{\dot{a}}\tilde{\lambda}_{3'}^{\dot{b}}-\lambda_{2'}^a\lambda_{1'}^b \tilde{\lambda}_{2'}^{\dot{a}}\tilde{\lambda}_{3'}^{\dot{b}}+\ldots\right)\frac{-y^3\la 12\ra}{\la 1'1\ra\la 2'2\ra} \nonumber \\
 &~~=\frac{i\pi \,y^4 \,s}{(16\pi^2)^2}\int d^5\vartheta \,\,\frac{e^{-i\phi}}{2\sqrt{6}}\left[c_{\theta_1}c_{\theta_3}\left(3+5c_{2\theta_2}-2 c_{2 \theta_1} s_{\theta_2}^2\right)+e^{-i\rho}\left(\ldots\right)\right]\!\tau^{ab\dot{a}\dot{b}}_{2,0}(\lambda_1,\lambda_2)\,\,\frac{e^{i\phi}}{c_{\theta_1}s_{\theta_2}^2 c_{\theta_3}} \nonumber \\
 &~~={\bf F}_0^{ab\dot{a}\dot{b}}(1_\varphi,2_\varphi)\,\,\frac{i\pi \,y^4}{(16\pi^2)^2}\,~ 2\int s_{\theta_1}c_{\theta_1}d\theta_1 \, s_{\theta_2}c_{\theta_2}d\theta_2 \left(3+5c_{2\theta_2}-2 c_{2 \theta_1} s_{\theta_2}^2\right) \nonumber \\
 &~~={\bf F}_0^{ab\dot{a}\dot{b}}(1_\varphi,2_\varphi)\,\,\frac{i\pi \,y^4}{(16\pi^2)^2}\, \left(\frac{3}{2}\right)~.
\end{align}
All other diagrams can be computed along these lines, though sometimes a different parametrization turns out more useful. For example, the following parametrization
\begin{align}\label{param1}
\lambda'_1&=\sqrt{1-\xi}\left( s_{\frac{\theta}{2}}e^{i\phi}\lambda_1 +  c_{\frac{\theta}{2}} \lambda_2 \right), \nonumber \\
\lambda'_2&=c_{\frac{\psi}{2}}\left( c_{\frac{\theta}{2}}\lambda_1 - e^{-i\phi} s_{\frac{\theta}{2}} \lambda_2\right)-s_{\frac{\psi}{2}}e^{-i\omega}\sqrt{\xi}\left( s_{\frac{\theta}{2}}e^{i\phi}\lambda_1 +  c_{\frac{\theta}{2}} \lambda_2 \right), \nonumber \\
\lambda'_3&=s_{\frac{\psi}{2}}e^{i\omega}\left( c_{\frac{\theta}{2}}\lambda_1 - e^{-i\phi} s_{\frac{\theta}{2}} \lambda_2 \right)+c_{\frac{\psi}{2}}\sqrt{\xi}\left( s_{\frac{\theta}{2}}e^{i\phi}\lambda_1 +  c_{\frac{\theta}{2}} \lambda_2 \right),
\end{align}
with measure
\be \label{eq:measure_PB}
\int d^5\vartheta \equiv 2 \, \int_{S^2} \frac{d\Omega(\theta,\phi)}{4\pi} \int_{S^2} \frac{d\Omega(\psi,\omega)}{4\pi} \int_0^1 d\xi (1-\xi)=1\,,
\ee
turns out to be more appropriate\footnote{What dictates the best choice of parametrization is the propagator structure of the diagram, since a given propagator can look more or less complicated for different choices of $\vartheta_i$.} to treat the following diagram
\begin{align}\label{3cutdiv}
    &\begin{tikzpicture}[line width=.7 pt, scale=0.6, baseline=(current bounding box.center)
    ]
	\draw (-1.2,0) to (-.1,1) ;
	\draw (-1.2,0) to (-.1,-1) ;
	\draw[dashed]  (-.6,.55) to  (-.1,.2) ;
	\draw[dotted] (0,1.4) -- (0,-1.4) ;
    \draw (.1,1) -- (1,.5) -- (1,-.5) -- (.1,-1) ;
    \draw[dashed] (.1,.2) to (.75,.65) ;
    \draw[dashed] (1,.5) to (2,1) ;
    \draw[dashed] (1,-.5) to (2,-1) ;
	\node at (-1.95,.2) {\footnotesize $T^{ab\dot{a}\dot{b}}$} ;
    \node at (2.2,-.8) {\tiny $1$} ;
    \node at (2.2,.8) {\tiny $2$} ;
    \node at (.4,-1.2) {\tiny $1'$} ;
    \node at (.4,0) {\tiny $3'$} ;
    \node at (.4,1.3) {\tiny $2'$} ;
 \end{tikzpicture}~~=\frac{i\pi \,s}{(16\pi^2)^2}\int d^5\vartheta\,\frac{-y}{2[3'2']}\left(\lambda_{1'}^a\lambda_{1'}^b \tilde{\lambda}_{1'}^{\dot{a}}\tilde{\lambda}_{3'}^{\dot{b}}-\lambda_{2'}^a\lambda_{1'}^b \tilde{\lambda}_{2'}^{\dot{a}}\tilde{\lambda}_{3'}^{\dot{b}}+\ldots\right)\frac{{-}y^3\la 13'\ra}{\la 1'1\ra\la 2'3'\ra} \nonumber \\
 &~~=\frac{i\pi \,y^4 \,s}{(16\pi^2)^2}\,\tau^{ab\dot{a}\dot{b}}_{2,0}\int d^5\vartheta ~(1-\xi)\left(\sqrt{\frac{2}{3}}\,d^2_{0,0}(\theta)c_{\frac{\psi}{2}}+\frac{e^{i(\phi-\omega)}}{\sqrt{\xi}}d_{0,-1}^2(\theta)s_{\frac{\psi}{2}}\right)\left(\frac{e^{i(\omega-\phi)}}{\sqrt{\xi}}t_{\frac{\theta}{2}}s_{\frac{\psi}{2}}-c_{\frac{\psi}{2}}\right) \nonumber \\
 &~~={\bf F}_0^{ab\dot{a}\dot{b}}(1_\varphi,2_\varphi)\,\,\frac{i\pi \,y^4}{(16\pi^2)^2}\times \lim_{\xi_0\to 0}\left(\,{2\ln\xi_0-3}+{\cal O}(\xi_0)\,\right) \rightarrow \infty\,.
 \end{align}
We see however that a naive evaluation gives an infinite yield, the origin of the divergence being the region where $\xi\to 0$, signaling that we need extra care when computing this diagram. A solution to this problem comes from noting that the two-loop diagram whose triple cut gives \eq{3cutdiv} -- which is $(a)$ in Fig.~\ref{2loopFeynman} -- has a one-loop subdivergence, whose counterterm admits a double cut and is associated to neither the left- nor the right-side amplitude (contrary, for example, to the counterterm that absorbs the one-loop subdivergence of diagram $(c)$ in Fig.~\ref{2loopFeynman}, which is clearly associated, after the two-cut, to ${\cal A}$). As a matter of fact, this counterterm diagram is precisely the one which, after being cut, exactly absorbs the divergence of \eq{3cutdiv}.

When \eq{3cutdiv} is treated in dimensional regularization consistently with the scheme and regularization choices used for computing ${\cal A}^{(1)}$ and ${\bf F}^{(1)}$ -- see Appendix~\ref{ddimex} -- one gets
\begin{equation}\label{divdiagrenorm}
    \left( \begin{tikzpicture}[line width=.7 pt, scale=0.6, baseline=(current bounding box.center)
    ]
	\draw (-1.2,0) to (-.1,1) ;
	\draw (-1.2,0) to (-.1,-1) ;
	\draw[dashed]  (-.6,.55) to  (-.1,.2) ;
	\draw[dotted] (0,1.4) -- (0,-1.6) ;
    \draw (.1,1) -- (1,.5) -- (1,-.5) -- (.1,-1) ;
    \draw[dashed] (.1,.2) to (.75,.65) ;
    \draw[dashed] (1,.5) to (2,1) ;
    \draw[dashed] (1,-.5) to (2,-1) ;
	\node at (-1.95,.2) {\footnotesize $T^{ab\dot{a}\dot{b}}$} ;
    \node at (2.2,-.8) {\tiny $1$} ;
    \node at (2.2,.8) {\tiny $2$} ;
    \node at (.4,-1.2) {\tiny $1'$} ;
    \node at (.4,0) {\tiny $3'$} ;
    \node at (.4,1.3) {\tiny $2'$} ;
 \end{tikzpicture}~~~ \right)_{\rm \!\!\! ren} = ~{\bf F}_0^{ab\dot{a}\dot{b}}(1_\varphi,2_\varphi)\,\,\frac{i\pi \,y^4}{(16\pi^2)^2}  \left( \frac{1}{2} \ln \frac{s}{\mu^2}  - \frac{7}{4} \right).
\end{equation}
All other diagrams turn out to be finite, the triple cut of Fig.~\ref{2loopFeynman}~$(f)$ being somewhat more challenging.
Summing together all of them, with symmetry factors when necessary (for example, both diagrams in \eq{3cutexample} and \eq{3cutdiv} come with a factor of 8, accounting for 2 helicity choices of the internal fermions, 2 distinct ways of attaching the internal scalar to the cut fermions and 2 possible ways in which the external scalars can be attached to the fermion loop), we obtain
\be
\sum_{\pm}  \, {\bf F}^{\mu\nu}\!\!\left(1'_{\psi_\pm}, 2'_{\psitil_\pm},3'_\varphi\right)\, S\!\left(1'_{\psi_\pm}, 2'_{\psitil_\pm},3'_\varphi\,;\,1_\varphi, 2_\varphi\right)={\bf F}_0^{\mu\nu}(1_\varphi,2_\varphi)\,\,\frac{i\pi \,y^4}{(16\pi^2)^2}  \left( 4 \ln\frac{s}{\mu^2}+8 \right).
\ee
Having all the ingredients, we can finally plug the result into \eq{masterYukawa} and read off the two-loop anomalous dimension.

\subsubsection{Extraction of two-loop anomalous dimension}

For illustrative purposes, in order to show the non-trivial ``bootstrapping'' at play, we consider all terms together instead of directly extracting the genuinely two-loop effects. Before doing that, we notice that, in a theory where (at least at one-loop order) $\gamma_{\rm cusp}=0$, we can simplify the exponent in \eq{master2}, thanks to the following identity
\be
a^*+\frac{i\pi}{2}\gamma_{\rm coll}^{(1)}={\rm Re}[a]=a_{\rm fin}^{(1)}-\frac{1}{2}\gamma_{\rm coll}^{(1)}\ln \frac{s}{\mu^2}\,.
\ee
Then, after dividing \eq{masterYukawa} by ${\bf F}^{\mu\nu}_0(1_\varphi,2_\varphi)$ and taking the logarithm of the equality, we finally get ($\ell\equiv y^2/16\pi^2$)
\begin{align}
  &\gamma^{(2)}_{\rm IR}|_\varphi+\gamma^{(1)}_{\rm coll}|_\varphi-\frac{2}{y}\,\left(a_{\rm fin}^{(1)}|_\varphi-\frac{1}{2}\gamma_{\rm coll}^{(1)}|_\varphi\ln \frac{s}{\mu^2}\right)\, \beta^{(1)}_y= \frac{1}{i\pi} \ln\left\lbrace 1+{i\pi\ell^2}\left(-\frac{94}{9}+i\pi\frac{4}{3}\right)\right. \nonumber\\
  &+\left. 2~ e^{\ell\left(-\frac{9}{2}+\frac{3}{2}\ln\frac{-s+i\varepsilon}{\mu^2}\right)}i\pi\left[ 2\ell+\ell^2\left(5\ln\frac{s}{\mu^2}-\frac{223}{9}+\frac{i\pi}{3}\right)\right]+{i\pi\ell^2}\left(4 \ln\frac{s}{\mu^2}+8 \right)\right\rbrace +{\cal O}(\ell^3) \nonumber \\
  &=\frac{1}{i\pi}\ln\left( 1+4(i\pi\ell)+ 8(i\pi\ell)^2  +20\, i\pi \ell^2 \ln\frac{s}{\mu^2}-70 \,i\pi\ell^2\right)+{\cal O}(\ell^3) \nonumber\\
  &=4\ell+ \left(20 \, \ln\frac{s}{\mu^2}- 70\right)\ell^2+{\cal O}(\ell^3)\,.
\end{align}
By comparing powers of $\ell$, we can see first of all that $\gamma^{(1)}_{\rm coll}|_\varphi=4(y^2/16\pi^2)$, consistently with what we found before and, thanks to this, that the non-local $\ln s$ piece cancel among the left- and right-hand sides. We can therefore extract the two-loop anomalous dimension, which is local (showing that IR divergences are of the collinear type in this sector and up to two loops) and given by
\be
\gamma^{(2)}_{\rm coll}|_\varphi=\frac{2}{y}\,a_{\rm fin}^{(1)}|_\varphi \,\beta^{(1)}_y -70\,\frac{y^4}{(16\pi^2)^2}=-10 \,\left(\frac{\,y^2\,}{16\pi^2}\right)^2\,,
\ee
which is twice the collinear anomalous dimension of particle $\varphi$, and also nicely agrees with the customary field anomalous dimension extracted from the (off-shell) two-point correlator of the $\varphi$ field \cite{Machacek:1983tz}. However we would like to stress that we did not just compute $\gamma_\varphi$ in a more complicated way: the point of this computation is to show what is the structure of IR divergences that one encounters in computing RG effects (especially operator anomalous dimensions) in a massless scheme and using \eq{chw}. Let us close by remarking once more how non-trivially the various pieces of the computation conspire to produce a real $\gamma_{\rm coll}$. Beyond the exact cancellation of the non-local pieces, we see that the real part in the argument of the log is exactly the square of the one-loop term (which is purely imaginary). This delicate game also shows the necessity of the complex conjugate in \eq{master}, starting from the two-loop order.

\subsection{Comments on Maximally Supersymmetric Yang-Mills}\label{SYM}

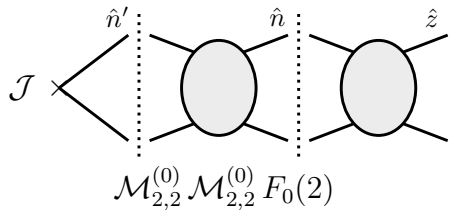
\begin{figure}[t]
\centering
\begin{tikzpicture}[line width=1.1 pt, scale=.7, baseline=(current bounding box.center)]

	\begin{scope}[shift={(5.4,-4)}]
	\draw (-1.5,0) to (-.2,1) ;
	\draw (-1.5,0) to (-.2,-1) ;
	\draw (1.5,.5) to  (.2,1) ;
	\draw (1.5,-.5) to  (.2,-1) ;
	\draw (1.5,.5) to  (2.8,1) ;
	\draw (1.5,-.5) to  (2.8,-1) ;
	\node at (-1.5,0) {$\times$} ;
	\node at (-2.2,0) {${\cal J}$} ;
	\draw[dotted] (0,1.4) -- (0,-1.4) ;
	\draw[fill=gray!15] (1.5,0) ellipse (.7 cm and .9 cm) ;
	\draw (4.5,.5) to  (3.2,1) ;
	\draw (4.5,-.5) to  (3.2,-1) ;
	\draw (4.5,.5) to  (5.8,1) ;
	\draw (4.5,-.5) to  (5.8,-1) ;
	\draw[dotted] (3,1.4) -- (3,-1.4) ;
	\draw[fill=gray!15] (4.5,0) ellipse (.7 cm and .9 cm) ;
	\node at (1.6,-1.9) {${\cal M}_{2,2}^{(0)}\, {\cal M}_{2,2}^{(0)} \,F_0(2)$} ;
	\node at (-0.4,1.35) {\footnotesize{$\hat{n}'$}} ;
	\node at (2.6,1.35) {\footnotesize{$\hat{n}$}} ;
	\node at (5.5,1.32) {\footnotesize{$\hat{z}$}} ;
	\end{scope}	
	
 \end{tikzpicture}
 \caption{\emph{Class of diagrams entering into the real part of \eq{master2} at two loops. On general grounds, they do not provide genuinely two-loop information, as they are $({\rm one-loop})^2$. These diagrams can have `cusp' divergences in each of the two cuts, requiring a double subtraction. See \eq{doubleint}.}}
 \label{fig:1loop^2}
 \end{figure}

Let us now consider another example which, from the point of view of \eq{master2}, is structurally simple. The maximally supersymmetric Yang-Mills theory in the limit $N_c\to\infty$ enjoys a number of properties, including (i) the fact that $\beta_g=0$ to all orders in perturbation theory, (ii) that there is only one species (or better multiplet) and finally (iii) that $\gamma_{\rm coll}^{(1)}=0$.\footnote{That $\gamma_{\rm coll}^{(1)}=0$ can be elegantly shown with the methods used in this paper, along the following lines. The free-theory limit of the minimal form factor of the energy-momentum tensor can be compactly expressed, with the supersymmetric on-shell formalism, as (see for example \cite{Arkani-Hamed:2008owk} for the notation)
\be\label{ffss}
{\bf F}_0^{ab\dot{a}\dot{b}}\!\left(\eta_1,\lambda_1,\tilde{\lambda}_1, \eta_2,\lambda_2,\tilde{\lambda}_2\right)=\frac{\varepsilon^{IJKL}}{12}\,(\eta_2\lambda_1-\eta_1\lambda_2)_I^a\,(\eta_2\lambda_1-\eta_1\lambda_2)_J^b\,(\eta_1\Tilde{\lambda}_1+\eta_2\Tilde{\lambda}_2)_K^{{\dot{a}}}\,(\eta_1\Tilde{\lambda}_1+\eta_2\Tilde{\lambda}_2)_L^{{\dot{b}}}\,.\nonumber 
\ee
Under a BCFW shift, with $\lambda'_1= \lambda_1+z \lambda_2$, $\Tilde{\lambda}'_2=\Tilde{\lambda}_2-z\Tilde{\lambda}_1$ and $\eta'_1= \eta_1+z\eta_2$, it can be seen that the form factor is unchanged (since each individual factor is). In order to compute $\gamma_{\rm IR}^{(1)}$, one has to convolute ${\bf F}_0$, as dictated by \eq{gammaIR1loop}, with the relevant $2\to 2$ tree-level amplitude, given in this case by the MHV amplitude
\be
{\cal A}=\frac{~\delta^{(8)}\!\!\left(\sum_{i=1}^4\tilde{\lambda}_i\eta_i\right)~}{\la 12\ra\la 23\ra \la 34 \ra\la 41\ra}\,, \nonumber
\ee
which decays as $1/z$ under a BCFW shift. As explained in \cite{Arkani-Hamed:2008owk}, the rational part of $\int {\bf F}_0(1',2'){\cal A}(1',2';1,2)$ -- which is, following the discussion at the end of Section~\ref{1loop}, what gives $\gamma_{\rm coll}$ at one loop -- is equal to the residue on the pole at infinite $z$ of the BCFW continuation of $\int {\bf F}_0(1',2'){\cal A}(1',2';1,2)$ (where $1'$ and $2'$ are shifted). Given the aforementioned behaviour of ${\bf F}_0$ and ${\cal A}$, we see that such a residue is zero, which concludes the proof.}
This means that the two-loop order in \eq{master2} simplifies to
\be
\left[i\pi \gamma_{\rm IR}^{(2)}-\frac{\pi^2}{2}\left(\gamma_{\rm IR}^{(1)}\right)^2\right]F_0(2)=i{\cal M}_{2,2}^{(1)}F_0(2)+i{\cal M}_{2,3}F(3)\,.
\ee
As explained in section 2.3 of \cite{Bern:2020ikv}, the two-loop equation is in general complex and can be further split into two. We get in this case
\begin{align}
\pi \gamma_{\rm IR}^{(2)}\,\,F_0(2)&={\rm Re} {\cal M}_{2,2}^{(1)} \,F_0(2)+{\cal M}_{2,3}F(3)\,,\label{real}\\
\frac{\pi^2}{2}\left(\gamma_{\rm IR}^{(1)}\right)^2 F_0(2)&={\rm Im} {\cal M}_{2,2}^{(1)} \,F_0(2)\,.\label{imaginary}
\end{align}
We see that the informative equation is the first one, as the second one contains only one-loop quantities. However, the consistency of the formalism requires it to be identically true. Since \eq{imaginary} provides a non-obvious identity, we study in detail how it happens to be true. The key observation is that, by virtue of the optical theorem, imaginary parts of loop amplitudes are computed by cuts. In formulas, we have
\be
{\rm Im} {\cal M}_{2,2}^{(1)}=\frac{1}{2}{\cal M}_{2,2}^{(0)}{\cal M}_{2,2}^{(0)}\,,
\ee
where, as usual, in the right-hand side of the equation a phase space integration over the intermediate state is understood, and the crucial factor 1/2 is dictated by the optical theorem \cite{Bern:2020ikv}. All in all, we find
\be\label{doublesub}
\left(\pi\gamma_{\rm IR}^{(1)}\right)^2 F_0(2)={\cal M}_{2,2}^{(0)}\, {\cal M}_{2,2}^{(0)} \,F_0(2)\,,
\ee
with a double phase space integration in the right-hand-side. How do we make sense of this equation? Remember that, on general grounds, the interpretation of the `cusp' contribution appearing in the dilation operator $D$ is in terms of a divergent integral, like in \eq{gammaIR}. What is the meaning of its square?
In light of the consistency of the master formula, we argue that it must be an integral being capable of absorbing the divergences of the right-hand-side of \eq{doublesub}, the divergences coming from the forward region of ${\cal M}_{2,2}^{(0)}$. If we parametrize with $\hat{n}$ ($\hat{n}'$) the direction of the two-particle state in the most external (internal) cut of Fig.~\ref{fig:1loop^2}, and we call $\hat{z}$ the direction of the external two-particle state (we are in the center of momentum frame of the system, where each pair of particles is back-to-back), we find that the divergences of the cut in Fig.~\ref{fig:1loop^2} can only be absorbed by an integral of the following form
\be\label{doubleint}
(\gamma_{\rm IR}^{(1)})^2=(\gamma_{\rm cusp}^{(1)})^2\int d{\hat{n}} \,\frac{1}{1-\hat{z}\cdot\hat{n}}\,\int d{\hat{n}'}\, \frac{1}{1-\hat{n}\cdot\hat{n}'}\,,
\ee
where, in the same notation, the divergent integral in \eq{gammaIR} would read (we set $\int d\hat{n}=2$)
\be
\gamma_{\rm IR}^{(1)}=\gamma_{\rm cusp}^{(1)}\int d{\hat{n}} \,\frac{1}{1-\hat{z}\cdot\hat{n}}\,.
\ee
In a general theory, the structure of the (one-loop)$^2$ cancellation can be more intricate, with contributions coming from $(\gamma_{\rm coll}^{(1)})^2$, that in super-Yang-Mills is zero, and also mixed ones.

\vspace{.5cm}

From a practical point of view, it is surely more interesting to say something about the structure of the genuinely two-loop contributions, i.e. \eq{real}, before concluding the section. \eq{gammaIR} is valid at all orders. Therefore we know that the right-hand-side of \eq{real} must necessarily take the form of a rational contribution, plus possibly a term with the same divergences as the phase space integral in the right-hand side of \eq{gammaIR}. Alternatively, if one prefers to dimensionally regulate the cuts, the result of the phase space integral must have the form of \eq{gammaIRdimreg}. This is quite a restrictive statement on the right-hand side of \eq{real}. Moreover, as we explained at the end of section \ref{1loop}, the two contributions to $\gamma_{\rm IR}$ (i.e. $\gamma_{\rm cusp}$ and $\gamma_{\rm coll}$) can be structurally distinguished, allowing one to project out the cusp contribution (as we said, this is especially useful if the focus is on the computation of UV divergences in the on-shell scheme). The concrete implementation of these ideas is deferred to future work.

\section{Conclusions}

\eq{chw} is an all order statement that, when expanded in loops, reveals to be extremely powerful, in that it allows to reconstruct the RG at $n$-loop order in terms of ($n-$1)-loop objects (in a sharply defined way). Like any computational technique, it requires practice to become operative. The main purpose of this work is to begin a discussion on the two-loop zoology of \eq{chw} when IR divergences are present, to show its consistence and disclose its simplicity.

Since UV and IR divergences are on the same ground in dimensional regularization (we consider here scaleless theories), one needs a way to disentangle them. For this purpose, we followed the path that was proposed in the literature of taking $\cal O$ to be conserved, to ensure that $\gamma_{\cal O}\equiv\gamma_{\rm UV}=0$ (strictly speaking, one has to also make sure that there are no identically conserved operators with the same quantum numbers of $\cal O$ \cite{Collins:2005nj}).

After some practice, one comes to agree with the experts on the statement that IR divergences are simple. In the language of \eq{chw}, this is essentially because they do not mix different form factors (up to color mixing in Yang-Mills). This property was the key for a simplification of the master formula at two-loops: in practice, for what concerns the two-particle cuts, we are instructed to compute only phase space integrals with the free theory (or tree-level) form factor $F_0$ (while, in general, one-loop $S$ matrix elements produce genuinely two-loop contributions).

We tried to let \eq{chw} talk. One (perhaps marginal, technical) suggestion that we captured is that we should better perform the phase space integrals in dimensional regularization, consistently with what we do for $S$ and $F$. This allows to make sense of some apparent ambiguities which are encountered in a straightforward interpretation of the master equation. See for example the discussion about \eq{gammaIRdimreg} and \eq{gammaIR}.

The method that we explored is completely general. This is a common feature in the on-shell approach, and the key to its versatility. Of course, some model-dependent property can make these techniques even more powerful, allowing them to be efficiently extended to higher loops (clearly, there is no intrinsic limitation in generalizing the formulas in this direction). 

Even though what was reported here is conceptually independent from the spinor-helicity formalism \cite{Cheung:2017pzi}, the practical implementation of the method greatly benefits from it. The connection between form factors and massive amplitudes is also illuminating and computationally useful \cite{Arkani-Hamed:2017jhn}.

Knowing the structure of IR divergences is necessary if one needs to subtract them, like in the study of the UV running of low energy EFTs in the on-shell massless scheme \cite{EliasMiro:2020tdv,Baratella:2020lzz,Baratella:2021guc,Jiang:2020mhe}. For what concerns the study of EFTs (and especially of the Standard Model EFT), a lot of recent activity was dedicated to the identification of amplitude bases for higher-dimensional operators \cite{Shadmi:2018xan,Ma:2019gtx,Aoude:2019tzn,Durieux:2019siw,AccettulliHuber:2021uoa}, and the study of various kinds of selection rules for mixing among operators, like helicity and angular momentum selection rules \cite{Cheung:2015aba,Craig:2019wmo,Jiang:2020rwz}. These sets of ideas, put together with the methods that were reported here, can be of great practical use for the systematic study of the RG at two loops \cite{Bern:2020ikv}.

The zoology of \eq{master2} is richer than what was reported in this work, but it surely is ``finite''. We leave for the future a more complete investigation of the formula, with its implications and simplifying strategies.

\vspace{.3cm}

\emph{Note added:} The present v2 substantially extends the work of v1, and has more authors. The additional content was originally planned to be part of a different paper.

\subsection*{Acknowledgements}
This research has been partially supported by the DFG Cluster of Excellence 2094 ORIGINS and the Collaborative Research Center SFB1258. The authors warmly thank, for the hospita-lity, the Munich Institute for Astro-, Particle and BioPhysics (MIAPbP), which is funded by the Deutsche Forschungsgemeinschaft (DFG, German Research Foundation) under Germany’s Excellence Strategy -- EXC-2094 -- 390783311. PB is greatly indebted to the Mainz Institute for Theoretical Physics (MITP) of the Cluster of Excellence PRISMA$^+$, for providing support and, especially, an extremely stimulating atmosphere during the workshop ``Amplitudes Meet BSM 2022''. This work was co-funded by the European Union (ERC, LoCoMotive, 101043686). Views and opinions expressed are however those of the authors only and do not necessarily reflect those of the European Union or the European Research Council.
Neither the European Union nor the granting authority can be held responsible for them. MS and TT would like to thank Lorenzo Tancredi and Cesare Mella for helpful discussions and an introduction to \texttt{Kira}.

\appendix

\section{$d$ - dimensional phase space integrals}\label{app:norm}

We want to derive the normalization of the phase space integrals appearing on the right-hand side of \eq{master} from first principles and for a generic number $d$ of space-time dimensions.

We start by fixing the norm of one-particle states, as
\be\label{normone}
\la \, {\bm q} \, | \, {\bm q'}\, \ra= 2 E_{\bm q} (2\pi)^{d-1} \delta^{(d-1)}({\bm q}- {\bm q'})\,.
\ee
Then we have
\be
_{\rm out}\la \,{\bm q'_1} \ldots {\bm q'_m}\,|\,{\bm q_1} \ldots {\bm q_n} \,\ra_{\rm in}=(2\pi)^d \delta^{(d)}(Q-Q')\,\, i {\cal A}({\bm q_1} \ldots {\bm q_n}\,;\, {\bm q'_1} \ldots {\bm q'_m})\,,
\ee
defining the phase and normalization of ${\cal A}$ on the right given the $S$-matrix element on the left (with multi-particle states normalized as in \eq{normone}), while ${ Q}={q}_1+\ldots +{ q}_n$.

\eq{normone} implies that the identity in the $n$-particle sector of the Fock space is given by
\be
\mathds{1}_{N=n}=\left(\prod_{i=1}^n \int\frac{d^{d-1}p_i}{2E_{i}\, (2\pi)^{d-1}}\right)|\, {\bm p}_1,{\bm p}_2 \ldots {\bm p}_n\,\ra \la \, {\bm p}_1,{\bm p}_2 \ldots  {\bm p}_n |\,,
\ee
assuming distinguishable particles. In the general case with identical particles, one should divide by $m!$ for each subset of $m$ identical particles in the multi-particle state.

Tracing the derivation of \eq{master0} in \cite{EliasMiro:2020tdv}, we can then fix the normalization of the phase space integrals appearing on its right-hand side. For example, the contributions to the right-hand side coming from $2\to n$ scattering amplitudes give
\begin{align}
e^{-i\pi { D}}F_{\cal O}^*( {\bm p_1}\ldots {\bm p_n} )&\supseteq \int \frac{d^{d-1}q_1}{2E_1(2\pi)^{d-1}} \int \frac{d^{d-1}q_2}{2E_2(2\pi)^{d-1}}(2\pi)^d \delta^{(d)}({ P}-{q_1}-{ q_2}) \nonumber \\
&~~~~~~~~~ \times ~F_{\cal O}^*( {\bm q_1},{\bm q_2} )~ i{\cal A}({\bm q_1},{\bm q_2}\,; \,{\bm p_1}\ldots {\bm p_n})\,.
\end{align}
It is convenient then to define a phase space measure for $n$-particle states that is appropriate for \eq{master0}, reading
\be
\int d\Phi_n^{(d)}\equiv \left(\prod_{i=1}^n \int\frac{d^d p_i}{(2\pi)^{d-1}}\delta^+(p_i^2-m_i^2)\right)(2\pi)^d\delta^{(d)}({ P}-{ p_1}-\ldots -{ p_n})\,.
\ee
Let us now study the phase space integrals relevant for the present work, with $n=2,3$ and $m_i=0$, calling $M=\sqrt{P^2}$. We emphasize the deviation from $d=4$, which may be relevant when dimensionally continuing the phase space integrals. For $n=2$, we find
\begin{align}
    \int d\Phi_2^{(d)}&=(2\pi)^{2-d}\int d^dp ~\delta^+({ p^2})\delta^+({ P^2}-2{ p}\cdot { P}) \nonumber \\
    &=(2\pi)^{2-d}\int d\Omega^{(d-2)}\int d|{\bm p}|\,|{\bm p}|^{d-2} \int dp_0 ~\delta^+(p_0^2-|{\bm p}|^2 )\delta^+(M^2-2Mp_0) \nonumber \\
    &=(2\pi)^{2-d} \frac{1}{2M}\int d\Omega^{(d-2)}\int d|{\bm p}|\,|{\bm p}|^{d-2}~\delta^+\left(|{\bm p}|^2-\left(\frac{M}{2}\right)^2\right) \nonumber \\
    &=(2\pi)^{2-d} \frac{1}{2M^2}\left(\frac{M}{2}\right)^{d-2} \int d\Omega^{(d-2)}=\frac{1}{32\pi^2}\left(\frac{M}{4\pi}\right)^{d-4} \int d\Omega^{(d-2)} \,.
\end{align}
We then have
\be
\int d\Omega^{(d-2)}=\frac{2\pi^{\frac{d-1}{2}}}{\Gamma\left(\frac{d-1}{2}\right)}=4\pi \left(e^{\gamma-2}4\pi\right)^{\frac{d-4}{2}}+{\cal O}\left((d-4)^2 \right)\,.
\ee
Plugging everything together, we finally find
\be
\int d\Phi_2^{(4-2\epsilon)}= \frac{1}{8\pi} \left(\frac{M^2 e^{\gamma-2}}{4\pi}\right)^{-\epsilon}+\,{\cal O}\left(\epsilon^2 \right)\,.
\ee
The three-particle phase space integral can be be written in terms of unconstrained variables as ($B$ is the Euler Beta function)
\begin{align}\label{3ind}
    \int d\Phi_3^{(d)}&= \frac{(2\pi)^{3-2d}}{64}\left(\frac{M}{2}\right)^{d-4}\int d\Omega^{(d-2)}_{1;23}\int d\Omega^{(d-2)}_{2;3} \int_0^{M^2} dm^2\left(\frac{m}{2}\right)^{d-4}\left(1-\frac{m^2}{M^2}\right)^{d-3} \nonumber \\
    &= \,M^2~\frac{(2\pi)^{3-2d}}{64}\left(\frac{M}{2}\right)^{2(d-4)} \left(\int d\Omega^{(d-2)}\right)^2 ~B\!\left[\frac{d}{2}-1,d-2\right] \nonumber \\
    &=\pi\,\frac{M^2}{\,(4\pi)^4} \left(\frac{M^2 e^{\gamma-\frac{13}{4}}}{4\pi}\right)^{-2\epsilon}+\,{\cal O}\left(\epsilon^2 \right)\,,
\end{align}
where $m^2=(p_2+p_3)^2$ is the invariant mass of the system of particles 2 and 3, while $\Omega_{1;23}$ is the direction of particle 1 in the rest frame of 1,2,3 and $\Omega_{2;3}$ is the direction of particle 2 in the rest frame of 2,3.

For general $n$, it can be shown that
\be
\int d\Phi_n^{(4)}\sim \pi \,\frac{M^{2n-4}}{(4\pi)^{2n-2}}~.
\ee
\subsection{An example from Yukawa theory}\label{ddimex}
Let us apply the parametrization of the three-particle phase space in \eq{3ind} to compute the diagram of \eq{3cutdiv}, divergent for $d\to 4$. We compute it with the standard loop technology (Feynman diagrams, gamma matrices, dimensional regularization, etc.). Since we are interested in the $J=2$ partial wave, we act on the relevant (cut) Feynman diagram with the $J=2$ projector (see Appendix~\ref{tautensors} for the definition of $\tau_{2,m}$)
\be
{\cal P}_{\,\rho\sigma}^{\,\mu\nu}=\,4\,\sum_m \,(-1)^m\,(\tau_{2,m})_{\rho\sigma}\,\tau_{2,-m}^{\,\mu\nu}\,,
\ee
and, tracing the arguments of Section~\ref{sec:3pPSI} and using \eq{scalarFtau}, we reduce to the computation of the following object
\begin{align}\label{divFeyn}
    \sum_{\sigma\,=\,\pm}\begin{tikzpicture}[line width=.7 pt, scale=0.6, baseline=(current bounding box.center)
    ]
	\draw (-1.2,0) to (-.1,1) ;
	\draw (-1.2,0) to (-.1,-1) ;
	\draw[dashed]  (-.6,.55) to  (-.1,.2) ;
	\draw[dotted] (0,1.4) -- (0,-1.4) ;
    \draw (.1,1) -- (1,.5) -- (1,-.5) -- (.1,-1) ;
    \draw[dashed] (.1,.2) to (.75,.65) ;
    \draw[dashed] (1,.5) to (2,1) ;
    \draw[dashed] (1,-.5) to (2,-1) ;
	\node at (-1.65,.2) {\footnotesize $T^{\mu\nu}$} ;
    \node at (2.2,-.8) {\tiny $1$} ;
    \node at (2.2,.8) {\tiny $2$} ;
    \node at (.4,-1.2) {\tiny $1'_\sigma$} ;
    \node at (.4,0) {\tiny $3'$} ;
    \node at (.4,1.3) {\tiny $2'_\sigma$} ;
 \end{tikzpicture}=\frac{i\sqrt{24}}{s}\,{\bf F}_0^{\mu\nu}(1_\varphi,2_\varphi)\,\tau_{2,0}^{\alpha\beta}(1,2)\int d\Phi^{(d)}_3 \frac{{\rm tr}\left[p'_1{\cal V}_{\alpha\beta}(p'_{23},p'_1)\,p'_{23}\,p'_2\,p'_{23}\,q\right]}{(p'^2_{23})^2q^2}\,,
\end{align}
where the trace is over a product of gamma matrices (we omit the Feynman slash), and
\be
{\cal V}_{\mu\nu}(k,k')=\frac{1}{4}\left[\,(k-k')_\mu \gamma_\nu+\gamma_\mu(k-k')_\nu -2 \,\eta_{\mu\nu}\,\gamma^{\alpha}(k-k')_\alpha\, \right]
\ee
is the vertex coming from the contraction of $T_{\mu\nu}$ with two Dirac-fermion legs, $p'_{23}=p'_2+p'_3$, $q=p'_1-p_1$, while
\be\label{taumunu20}
\tau_{2,0}^{\alpha\beta}(1,2)=\frac{\la 1 |\sigma^\alpha | 1 ]\la 1 |\sigma^\beta | 1 ]-\la 1 |\sigma^\alpha | 1 ]\la 2 |\sigma^\beta | 2 ]-\la 1 |\sigma^\alpha | 2 ]\la 2 |\sigma^\beta | 1 ]}{4\sqrt{6}\la 12\ra [21]}+(1\leftrightarrow 2)\,.
\ee
The tedious algebraic steps that follow, which we skip, can be simplified by exploiting the fact that $\tau_{2,0}^{\mu\nu}\eta_{\mu\nu}=\tau_{2,0}^{\mu\nu}P_\mu=0$ and making use of the following phase space identities
\begin{align}
    \int d\Omega^{(d-2)}_{2;3} p'^\mu_2 &=\frac{1}{2}\, p'^\mu_{23} \int d\Omega^{(d-2)}_{2;3}\,, \\
        \int d\Omega^{(d-2)}_{1;23} p'^\mu_1 &=\frac{1}{2}\left( 1-\frac{m^2}{M^2}\right)\, P^\mu \int d\Omega^{(d-2)}_{1;23}\,, \\
    \int d\Omega^{(d-2)}_{1;23} p'^\mu_1 p'^\nu_1 &=\frac{d}{4(d-1)}\left( 1-\frac{m^2}{M^2} \right)^2 \left(P^\mu P^\nu-\frac{1}{d}\,\eta^{\mu\nu}P^2\right)\int d\Omega^{(d-2)}_{1;23}\,,
\end{align}
and one finally arrives at
\begin{align}
   \sum_{\sigma\,=\,\pm}\begin{tikzpicture}[line width=.7 pt, scale=0.6, baseline=(current bounding box.center)
    ]
	\draw (-1.2,0) to (-.1,1) ;
	\draw (-1.2,0) to (-.1,-1) ;
	\draw[dashed]  (-.6,.55) to  (-.1,.2) ;
	\draw[dotted] (0,1.4) -- (0,-1.4) ;
    \draw (.1,1) -- (1,.5) -- (1,-.5) -- (.1,-1) ;
    \draw[dashed] (.1,.2) to (.75,.65) ;
    \draw[dashed] (1,.5) to (2,1) ;
    \draw[dashed] (1,-.5) to (2,-1) ;
	\node at (-1.65,.2) {\footnotesize $T^{\mu\nu}$} ;
    \node at (2.2,-.8) {\tiny $1$} ;
    \node at (2.2,.8) {\tiny $2$} ;
    \node at (.4,-1.2) {\tiny $1'_\sigma$} ;
    \node at (.4,0) {\tiny $3'$} ;
    \node at (.4,1.3) {\tiny $2'_\sigma$} ;
 \end{tikzpicture}&= {\bf F}_0^{\mu\nu}(1_\varphi,2_\varphi)\,\frac{i \,y^4}{\,512\,}\,(2\pi)^{3-2d} \left(\frac{2\mu}{M}\right)^{4\epsilon} \left(\int d\Omega^{(d-2)}\right)^2c_d~~~~~ \\
    \times \int_0^1 d\xi ~\xi^{\frac{d}{2}-3}\,(1-\xi)^{d-2}&\,\frac{(d-4)\xi+3}{d-1} 
    ={\bf F}_0^{\mu\nu}(1_\varphi,2_\varphi)\,\frac{i\pi\,y^4}{(16\pi^2)^2}\left(-\frac{1}{\epsilon} -\frac{37}{6}+\frac{c'}{2}+2 \ln\frac{e^\gamma M^2}{4\pi\mu^2}\right), \nonumber
\end{align}
where $\xi=m^2/M^2$ is the $d$-dimensional analogue of the parameter entering \eq{param1}, and we called $c_d={\rm tr}[\mathds{1}]=4-2c'\epsilon+{\cal O}(\epsilon^2)$. Thanks to the usual theorems on the cancellation of subdivergences, the above divergence should be cancelled by the following expression
\begin{align}
   \sum_{\psi,\bar{\psi}}\begin{tikzpicture}[line width=.7 pt, scale=0.6, baseline=(current bounding box.center)
    ]
	\draw (-1.2,0) to (-.1,1) ;
	\draw (-1.2,0) to (-.1,-1) ;
	\draw[dotted] (0,1.4) -- (0,-1.4) ;
    \draw (.1,1) -- (1,.5) -- (1,-.5) -- (.1,-1) ;
    \draw[dashed] (1,.5) to (2,1) ;
    \draw[dashed] (1,-.5) to (2,-1) ;
	\node at (-1.65,.2) {\footnotesize $T^{\mu\nu}$} ;
    \node at (2.2,-.8) {\tiny $1$} ;
    \node at (2.2,.8) {\tiny $2$} ;
    \node at (.4,-1.2) {\tiny $1'_-$} ;
    \node at (.4,1.3) {\tiny $2'_+$} ;
 \end{tikzpicture} \times ~\frac{\gamma_{\rm coll}^{(1)}|_\psi}{2\epsilon}~={\bf F}_0^{\mu\nu}(1_\varphi,2_\varphi)\,\frac{i\pi\,y^4}{(16\pi^2)^2}\left(\frac{1}{\epsilon} +\frac{8}{3}-\frac{c'}{2}- \ln\frac{e^\gamma M^2}{4\pi\mu^2}\right).
\end{align}
Adding up the two expressions and reabsorbing the factor of $4\pi e^{-\gamma}$ via a redefinition of $
\mu^2$ (consistently with the $\overline{\rm MS}$ scheme), we find that the divergence goes away along with the dependence on the choice of $c_d$, and we get precisely twice as \eq{divdiagrenorm}, because we considered both polarizations together, as incorporated in the numerator of the fermionic propagators.

\section{Covariant $J=2$ tensors}\label{tautensors}

The following tensors are constructed out of two arbitrary reference spinors $\zeta$ and $\nu$ -- provided they are not parallel -- and their complex conjugated spinors $\Tilde{\zeta}$ and $\Tilde{\nu}$. They are of the form $\tau^{a  b \dot{a}\dot{b}}_{2,m}$, and read
\begin{align}
\tau_{2,2}(\zeta,\nu)&=\frac{\zeta\zeta\tilde{\nu}\tilde{\nu}}{\la \zeta\nu\ra[\nu\zeta]}\,,\\
\tau_{2,1}(\zeta,\nu)&=\frac{1}{2\,\la \zeta\nu\ra[\nu\zeta]}\left(-\zeta\zeta\tilde{\nu}\tilde{\zeta}-\zeta\zeta\tilde{\zeta}\tilde{\nu}+\zeta\nu\tilde{\nu}\tilde{\nu}+\nu\zeta\tilde{\nu}\tilde{\nu}\right)\,,\\
\tau_{2,0}(\zeta,\nu)&=\frac{1}{\sqrt{6}\,\la \zeta\nu\ra[\nu\zeta]}\left(\zeta\zeta\tilde{\zeta}\tilde{\zeta}-\zeta\nu\tilde{\nu}\tilde{\zeta}-\zeta\nu\tilde{\zeta}\tilde{\nu}-\nu\zeta\tilde{\nu}\tilde{\zeta}-\nu\zeta\tilde{\zeta}\tilde{\nu}+\nu\nu\tilde{\nu}\tilde{\nu}\right)\,,\\
\tau_{2,-1}(\zeta,\nu)&=\frac{1}{2\,\la \zeta\nu\ra[\nu\zeta]}\left(\zeta\nu\tilde{\zeta}\tilde{\zeta}+\nu\zeta\tilde{\zeta}\tilde{\zeta}-\nu\nu\tilde{\nu}\tilde{\zeta}-\nu\nu\tilde{\zeta}\tilde{\nu}\right)\,,\\
\tau_{2,-2}(\zeta,\nu)&=\frac{\nu\nu\tilde{\zeta}\tilde{\zeta}}{\la \zeta\nu\ra[\nu\zeta]}\,.
\end{align}
They have simple contractions
\be
(\tau_{2,m})_{a  b \dot{a}\dot{b}}\, (\tau_{2,-m'})^{a  b \dot{a}\dot{b}}=(-1)^m \,\delta_{mm'}\,,
\ee
and form a $J=2$ multiplet of $SO(3)$. In fact the whole multiplet can be generated from $\tau_{2,2}$ by acting with the lowering operator in the standard way, with ${\bf J_-}|\zeta\ra=|\nu\ra$ and ${\bf J_-} |{\nu}]=-|{\zeta}]$.
We can also define the related tensors
\be
\tau_{2,m}^{\mu\nu}\equiv \frac{1}{4}\,\sigma^\mu_{a \dot{a}}\,\sigma^\nu_{b \dot{b}}\,\tau^{a  b \dot{a}\dot{b}}_{2,m}\,,
\ee
where the factor $1/4\,$ is chosen to be consistent with the standard definition $p_{a \dot{a}}\equiv p_\mu\sigma^\mu_{a \dot{a}}$, which implies that $p_\mu=\tfrac{1}{2}\,\overline{\sigma}_\mu^{\dot{b} b}p_{b \dot{b}}$. For example, we have
\be
\tau_{2,2}^{\mu\rho}(\zeta,\nu)=\frac{\,\la \zeta | \sigma^\mu |\nu]\,\la \zeta | \sigma^\rho |\nu]\,}{4\,\la \zeta\nu\ra[\nu\zeta]}\,,
\ee
or \eq{taumunu20}. These two-index tensors satisfy a similar orthogonality condition as their four-index counterparts
\be
(\tau_{2,m})_{\mu\nu}\,(\tau_{2,-m'})^{\mu\nu}=\frac{1}{4}\,(-1)^m\, \delta_{mm'}\,.
\ee
The following relation involving the Wigner $D$-matrix holds
\be\label{tauangular}
\tau_{2,m'}(\lambda'_1,\lambda'_2)=\sum_m \,e^{i(m-m')\phi}\, d^2_{m,m'}(\theta)\,\tau_{2,m}(\lambda_1,\lambda_2)=\sum_m \,{D}^2_{m,m'}(\phi,\theta,-\phi)\, \tau_{2,m}(\lambda_1,\lambda_2)\,,
\ee
where the relation among the primed and unprimed variables is given in \eq{lambdaprime}.

\section{Complete two-loop form factor in Yukawa theory}\label{appC}

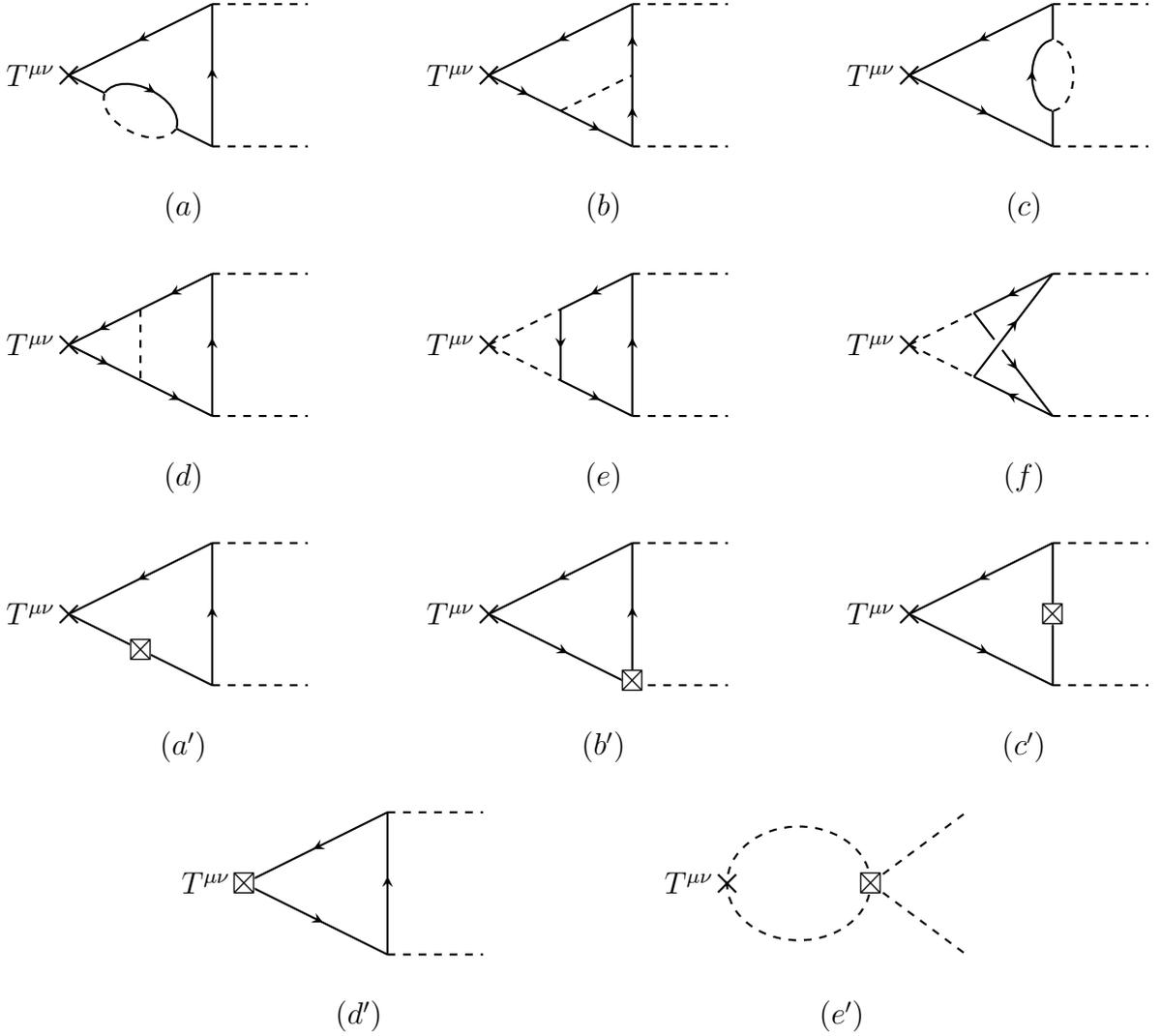
\begin{figure}
 \hspace*{\fill}
\begin{minipage}{.3\linewidth}
\begin{tikzpicture}[baseline={(current bounding box.center)}, scale=1.3]
    \draw[fermionbar,thick, >=stealth] (0,0) to (1.5,.75) ;
    \draw[white] (0,0) to coordinate[pos=0.25] (I1) coordinate[pos=0.75] (I2) (1.5,-.75) ;
    \draw[thick] (0,0) to (I1) ;
    \draw[thick] (I2) to (1.5,-.75) ;
    \draw[scalarnoarrow,thick, >=stealth] (1.5,-.75) to (2.5,-.75) ;
    \draw[scalarnoarrow,thick, >=stealth] (1.5,.75) to (2.5,.75) ;
    \draw[fermion,thick, >=stealth] (1.5,-.75) to (1.5,.75) ;
    \draw[fermion,thick, >=stealth] (I1) to[bend left=80] (I2) ;
    \draw[scalarnoarrow,thick, >=stealth] (I1) to[bend right=80] (I2) ;
    \draw[white] (I1) to (I2) ;
    \draw (0,0) node[cross out, thick, draw, inner sep=0pt, outer sep=3pt] {} ;
	\node at (-0.3,0) {$T^{\mu\nu} ~\;$} ;
     \node at (1.2,-1.4) {$(a)$} ;
 \end{tikzpicture}
 \vspace{.5cm}
 \end{minipage}
 \hspace*{\fill}
 \begin{minipage}{.3\linewidth}
 \begin{tikzpicture}[baseline={(current bounding box.center)}, scale=1.3]
    \draw[fermionbar,thick, >=stealth] (0,0) to (1.5,.75) ;
    \draw[white] (0,0) to coordinate[pos=0.5] (I1) (1.5,-.75) ;
    \draw[fermion,thick, >=stealth] (0,0) to (I1);
    \draw[fermion,thick, >=stealth] (I1) to (1.5,-.75);
    \draw[scalarnoarrow,thick] (1.5,-.75) to (2.5,-.75) ;
    \draw[scalarnoarrow,thick] (1.5,.75) to (2.5,.75) ;
    \draw[white] (1.5,-.75) to coordinate[pos=0.5] (I2) (1.5,.75) ;
    \draw[fermion,thick, >=stealth] (1.5,-.75) to (I2);
    \draw[fermion,thick, >=stealth] (I2) to (1.5,.75);
    \draw[scalarnoarrow, thick] (I1) to (I2);
    \draw (0,0) node[cross out, thick, draw, inner sep=0pt, outer sep=3pt] {} ;
	\node at (-0.3,0) {$T^{\mu\nu}~ \;$} ;
    \node at (1.2,-1.4) {$(b)$} ;
 \end{tikzpicture}
 \vspace{.5cm}
 \end{minipage}
\hspace*{\fill}
 \begin{minipage}{.3\linewidth}
  \begin{tikzpicture}[baseline={(current bounding box.center)}, scale=1.3]
    \draw[fermionbar,thick, >=stealth] (0,0) to (1.5,.75) ;
    \draw[fermion,thick, >=stealth] (0,0) to (1.5,-.75) ;
    \draw[scalarnoarrow,thick, >=stealth] (1.5,-.75) to (2.5,-.75) ;
    \draw[scalarnoarrow,thick, >=stealth] (1.5,.75) to (2.5,.75) ;
    \draw[white] (1.5,-.75) to coordinate[pos=0.25] (I1) coordinate[pos=0.75] (I2) (1.5,.75) ;
    \draw[thick] (1.5,-.75) to (I1) ;
    \draw[thick] (I2) to (1.5,.75) ;
    \draw[fermion,thick, >=stealth] (I1) to[bend left=80] (I2) ;
    \draw[scalarnoarrow,thick, >=stealth] (I1) to[bend right=80] (I2) ;
    \draw[white] (I1) to (I2) ;
    \draw (0,0) node[cross out, thick, draw, inner sep=0pt, outer sep=3pt] {} ;
	\node at (-0.3,0) {$T^{\mu\nu}~ \;$} ;
    \node at (1.2,-1.4) {$(c)$} ;
 \end{tikzpicture}
 \vspace{.5cm}
 \end{minipage}
 \hspace*{\fill}
 \begin{minipage}{.3\linewidth}
  \begin{tikzpicture}[baseline={(current bounding box.center)}, scale=1.3]
    \draw[white] (0,0) to coordinate[pos=0.5] (I1) (1.5,.75) ;
    \draw[fermionbar,thick, >=stealth] (0,0) to (I1);
    \draw[fermionbar,thick, >=stealth] (I1) to (1.5,.75);
    \draw[white] (0,0) to coordinate[pos=0.5] (I2) (1.5,-.75) ;
    \draw[fermion,thick, >=stealth] (0,0) to (I2);
    \draw[fermion,thick, >=stealth] (I2) to (1.5,-.75);
    \draw[scalarnoarrow, thick] (I1) to (I2);
    \draw[scalarnoarrow,thick] (1.5,-.75) to (2.5,-.75) ;
    \draw[scalarnoarrow,thick] (1.5,.75) to (2.5,.75) ;
    \draw[fermion, thick, >=stealth] (1.5,-.75) to (1.5,.75) ;
    \draw (0,0) node[cross out, thick, draw, inner sep=0pt, outer sep=3pt] {} ;
	\node at (-0.3,0) {$T^{\mu\nu}~ \;$} ;
    \node at (1.2,-1.4) {$(d)$} ;
 \end{tikzpicture}
 \vspace{.5cm}
\end{minipage}
\hspace*{\fill}
\begin{minipage}{.3\linewidth}
  \begin{tikzpicture}[baseline={(current bounding box.center)}, scale=1.3]
    \draw[white] (0,0) to coordinate[pos=0.5] (I1) (1.5,.75) ;
    \draw[scalarnoarrow,thick] (0,0) to (I1);
    \draw[fermionbar,thick, >=stealth] (I1) to (1.5,.75);
    \draw[white] (0,0) to coordinate[pos=0.5] (I2) (1.5,-.75) ;
    \draw[scalarnoarrow,thick] (0,0) to (I2);
    \draw[fermion,thick, >=stealth] (I2) to (1.5,-.75);
    \draw[fermion, thick, >=stealth] (I1) to (I2);
    \draw[scalarnoarrow,thick] (1.5,-.75) to (2.5,-.75) ;
    \draw[scalarnoarrow,thick] (1.5,.75) to (2.5,.75) ;
    \draw[fermion, thick, >=stealth] (1.5,-.75) to (1.5,.75) ;
    \draw (0,0) node[cross out, thick, draw, inner sep=0pt, outer sep=3pt] {} ;
	\node at (-0.3,0) {$T^{\mu\nu}~ \;$} ;
    \node at (1.2,-1.4) {$(e)$} ;
 \end{tikzpicture}
 \vspace{.5cm}
\end{minipage}
\hspace*{\fill}
 \begin{minipage}{.3\linewidth}
   \begin{tikzpicture}[baseline={(current bounding box.center)}, scale=1.3]
    \draw[white] (0,0) to coordinate[pos=0.45] (I1) (1.5,.75) ;
    \draw[scalarnoarrow,thick] (0,0) to (I1);
    \draw[fermionbar,thick, >=stealth] (I1) to (1.5,.75);
    \draw[white] (0,0) to coordinate[pos=0.45] (I2) (1.5,-.75) ;
    \draw[scalarnoarrow,thick] (0,0) to (I2);
    \draw[fermionbar,thick, >=stealth] (I2) to (1.5,-.75);
    \draw[fermion, thick, >=stealth] (I1) to coordinate[pos=0.26] (O1) coordinate[pos=0.36] (O2) (1.5,-.75);
    \draw[white, opacity=1, ultra thick] (O1) to (O2);
    \draw[fermion, thick, >=stealth] (I2) to (1.5,.75);
    \draw[scalarnoarrow,thick] (1.5,-.75) to (2.5,-.75) ;
    \draw[scalarnoarrow,thick] (1.5,.75) to (2.5,.75) ;
    \draw (0,0) node[cross out, thick, draw, inner sep=0pt, outer sep=3pt] {} ;
	\node at (-0.3,0) {$T^{\mu\nu} ~\;$} ;
    \node at (1.2,-1.4) {$(f)$} ;
 \end{tikzpicture}
   \vspace{.5cm}
 \end{minipage}
   \hspace*{\fill}
 \begin{minipage}{.3\linewidth}
  \begin{tikzpicture}[baseline={(current bounding box.center)}, scale=1.3]
    \draw[fermionbar,thick, >=stealth] (0,0) to (1.5,.75) ;
    \draw[thick] (0,0) to (1.5,-.75) ;
    \draw[fermion,thick, >=stealth] (1.5,-.75) to (1.5,.75) ;
    \draw[scalarnoarrow,thick, >=stealth] (1.5,-.75) to (2.5,-.75) ;
    \draw[scalarnoarrow,thick, >=stealth] (1.5,.75) to (2.5,.75) ;
    \draw (0,0) node[cross out, thick, draw, inner sep=0pt, outer sep=3pt] {} ;
	\node at (-0.3,0) {$T^{\mu\nu}~ \;$} ;
    \node[rectangle,draw=white,fill=white] (r) at (.75,-.375){};
    \node at (.75,-.375) {$\boxtimes$} ;
    \node at (1.2,-1.4) {$(a')$} ;
 \end{tikzpicture}
     \vspace{.5cm}
  \end{minipage}
  \hspace*{\fill}
\begin{minipage}{.3\linewidth}
  \begin{tikzpicture}[baseline={(current bounding box.center)}, scale=1.3]
    \draw[fermionbar,thick, >=stealth] (0,0) to (1.5,.75) ;
    \draw[fermion,thick, >=stealth] (0,0) to (1.5,-.75) ;
    \draw[fermion,thick, >=stealth] (1.5,-.75) to (1.5,.75) ;
    \draw[scalarnoarrow,thick, >=stealth] (1.5,-.75) to (2.5,-.75) ;
    \draw[scalarnoarrow,thick, >=stealth] (1.5,.75) to (2.5,.75) ;
    \draw (0,0) node[cross out, thick, draw, inner sep=0pt, outer sep=3pt] {} ;
    \node[rectangle,draw=white,fill=white] (r) at (1.5,-.7){};
    \node at (1.5,-.7) {$\boxtimes$} ;
	\node at (-0.3,0) {$T^{\mu\nu}~ \;$} ;
    \node at (1.2,-1.4) {$(b')$} ;
 \end{tikzpicture}
     \vspace{.5cm}
  \end{minipage}
 \hspace*{\fill}
 \begin{minipage}{.3\linewidth}
  \begin{tikzpicture}[baseline={(current bounding box.center)}, scale=1.3]
    \draw[fermionbar,thick, >=stealth] (0,0) to (1.5,.75) ;
    \draw[fermion,thick, >=stealth] (0,0) to (1.5,-.75) ;
    \draw[thick] (1.5,-.75) to (1.5,.75) ;
    \draw[scalarnoarrow,thick, >=stealth] (1.5,-.75) to (2.5,-.75) ;
    \draw[scalarnoarrow,thick, >=stealth] (1.5,.75) to (2.5,.75) ;
    \draw (0,0) node[cross out, thick, draw, inner sep=0pt, outer sep=3pt] {} ;
    \node[rectangle,draw=white,fill=white] (r) at (1.5,0){};
    \node at (1.5,0) {$\boxtimes$} ;
	\node at (-0.3,0) {$T^{\mu\nu}~ \;$} ;
    \node at (1.2,-1.4) {$(c')$} ;
 \end{tikzpicture}
    \vspace{.5cm}
  \end{minipage}
    \hspace*{0.16\linewidth}
  \begin{minipage}{0.25\linewidth}
 \centering
  \begin{tikzpicture}[baseline={(current bounding box.center)}, scale=1.3]
    \draw[fermionbar,thick, >=stealth] (0,0) to (1.5,.75) ;
    \draw[fermion,thick, >=stealth] (0,0) to (1.5,-.75) ;
    \draw[fermion, thick, >=stealth] (1.5,-.75) to (1.5,.75) ;
    \draw[scalarnoarrow,thick, >=stealth] (1.5,-.75) to (2.5,-.75) ;
    \draw[scalarnoarrow,thick, >=stealth] (1.5,.75) to (2.5,.75) ;
    \draw (0,0) node[cross out, thick, draw, inner sep=0pt, outer sep=3pt] {} ;
    \node[rectangle,draw=white,fill=white] (r) at (0,0){};
    \node at (0,0) {$\boxtimes$} ;
	\node at (-0.3,0) {$T^{\mu\nu}~ \;$} ;
    \node at (1.2,-1.4) {$(d')$} ;
 \end{tikzpicture}
     \vspace{.5cm}
  \end{minipage}
   \hspace*{0.12\linewidth}
    \begin{minipage}{0.25\linewidth}
 \centering
  \begin{tikzpicture}[baseline={(current bounding box.center)}, scale=1.3]
    \draw[scalarnoarrow, thick, fill=white] (0.75,0) ellipse (0.75 and 0.6) ;
    \draw[scalarnoarrow,thick] (1.5,0) to (2.5,-.75) ;
    \draw[scalarnoarrow,thick] (1.5,0) to (2.5,.75) ;
    \draw (0,0) node[cross out, thick, draw, inner sep=0pt, outer sep=3pt] {} ;
    \node[rectangle,draw=white,fill=white] (r) at (1.5,0){};
    \node at (1.5,0) {$\boxtimes$} ;
	\node at (-0.3,0) {$T^{\mu\nu}~ \;$} ;
    \node at (1.2,-1.4) {$(e')$} ;
 \end{tikzpicture}
     \vspace{.5cm}
  \end{minipage}
 \caption{\emph{$(a)$--$\,(f)$ Two-loop Feynman diagrams, up to symmetry operations, entering the computation of the two-loop corrections to the form factor ${\bf F}^{\mu\nu}(1_{\varphi}, 2_{\varphi})$, whose expression up to the one-loop order is given in \eq{formfactors}. $(a')$--$\,(e')$ At the same order in perturbation theory, there are several one-loop diagrams with insertions of one-loop counterterms, which are shown in the last two lines of the figure. As explained in Sec.~\ref{sec:yukawa}, the double cut of $(a')$, which carries a divergence due to the $\boxtimes$ insertion, precisely cancels the divergence of the three-particle phase space integral associated to the triple cut of $(a)$, given by \eq{3cutdiv}. The counterterm diagrams $(b')$ and $(c')$ shown in the picture are responsible -- once cut -- for the renormalization of ${\cal A}({\psi},{\psi} ; \varphi, \varphi)$ in \eq{fermionscalar}, while diagrams $(d')$ and $(e')$ renormalize the form factor ${\bf F}^{\mu\nu}(\psi,\psi)$ in \eq{formfactorf} and ${\cal A}(\varphi, \varphi ; \varphi, \varphi)$ in \eq{scalar1loop}, respectively.}}\label{2loopFeynman}
 \end{figure}

 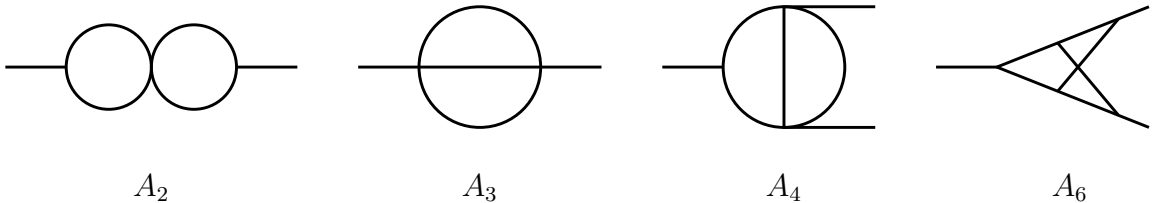
\begin{figure}[t]
\centering
\begin{tikzpicture}[line width=1.1 pt, scale=.8, baseline=(current bounding box.center)]
\draw (3.8,0) to (7.8,0);
\draw (5.8,0) circle (1cm);
\node at (5.8,-2) {$A_3$};
\draw (-2,0) to (-1,0) ;
\draw (-.3,0) circle (.7cm);
\draw (1.1,0) circle (.7cm);
\draw (1.8,0) to (2.8,0) ;
\node at (.4,-2) {$A_2$};
\draw (8.8,0) to (9.8,0);
\draw (10.8,0) circle (1cm);
\draw (12.3,1) -- (10.8,1) -- (10.8,-1) -- (12.3,-1);
\node at (10.8,-2) {$A_4$};
\draw (13.3,0) to (14.3,0);
\draw (16.8,1) -- (14.3,0) -- (16.8,-1);
\draw (15.3,.4) -- (16.3,-.8);
\draw (15.3,-.4) -- (16.3,.8);
\node at (15.5,-2) {$A_6$};
 \end{tikzpicture}
 \caption{\emph{Scalar integrals relevant for the computation in Appendix~\ref{appC}. We adopted a standard notation for the integrals, whose analytic expressions in $d$ dimensions can be found e.g. in \cite{Gehrmann:2005pd,Gehrmann:1999as}. }}
 \label{fig:scalarbasis}
 \end{figure}

The main purpose of this Appendix is to cross check the results of Section~\ref{sec:yukawa} using traditional methods, i.e. with Feynman diagrams techniques. The infrared anomalous dimension is extracted from the appropriate renormalization constant $Z$ as detailed below.

We regulate the theory by going to $d$ spacetime dimensions, where the non-interacting part of the traceless and symmetric energy-momentum tensor in the scalar sector is given by 
\begin{equation}
    T_{\mu\nu}|_{\varphi,0}=\partial_\mu\varphi\partial_\nu\varphi-\frac{1}{2}\eta_{\mu\nu}\partial_\alpha\varphi\partial^\alpha\varphi+\frac{d-2}{4(d-1)}\left(\partial_\mu\partial_\nu-\eta_{\mu\nu}\partial^2\right)\varphi^2\,,
\end{equation}
where we kept the last term to depend explicitly on $d$, such that $\eta^{\mu\nu}T_{\mu\nu}=0$ in all dimensions. In $d=4$ it reduces to the usual factor of $1/6$. The form factor that we are going to compute is given by
\be\label{yukawaff}
  _{\rm out}\langle \,1_\varphi, 2_\varphi \,| \, 
{\tau}_{2,0}^{\mu\nu}(\lambda_1,\lambda_2)\,T_{\mu\nu}(x) \, |\, 0\, \rangle \equiv e^{i (p_1+p_2)\cdot x} \,\,{\bf F}(1_\varphi, 2_\varphi ),
\ee
where ${\tau}_{2,0}$ is defined in Appendix~\ref{tautensors} and given explicitly in \eq{taumunu20}. 
After contracting our Feynman diagrams -- we show in Fig.~\ref{2loopFeynman} those that arise at two loops -- with ${\tau}_{2,0}$, the numerators of the so obtained integrals only feature scalar products of external and loop momenta. Our expression can then, by the use of standard techniques, be reduced to scalar integrals (i.e. with unit numerators) by rewriting the scalar products in the numerators in terms of inverse propagators such that they cancel parts of the denominator. The resulting scalar integrals can be further reduced by using Integration-By-Parts and Lorentz-Invariance identities. In practice, we use the software \texttt{Kira} \cite{Maierhofer:2017gsa}, which fully automatizes the reduction of scalar integrals to the smaller basis. We find that, in our case, all relevant quantities can be expanded over the four objects of
Fig.~\ref{fig:scalarbasis}.\\

Having an expansion over a standard basis can be useful for many purposes, for example as a powerful means to compute unitarity cuts. We want to illustrate this using a simple example. Schematically, we can decompose
\begin{align}
     \begin{tikzpicture}[line width=.7 pt, scale=0.8, baseline=(current bounding box.center)
    ]
    \draw[fermionbar,thick, >=stealth] (0,0) to (1.5,.75) ;
    \draw[white] (0,0) to coordinate[pos=0.5] (I1) (1.5,-.75) ;
    \draw[fermion,thick, >=stealth] (0,0) to (I1);
    \draw[fermion,thick, >=stealth] (I1) to (1.5,-.75);
    \draw[scalarnoarrow,thick] (1.5,-.75) to (2.5,-.75) ;
    \draw[scalarnoarrow,thick] (1.5,.75) to (2.5,.75) ;
    \draw[white] (1.5,-.75) to coordinate[pos=0.5] (I2) (1.5,.75) ;
    \draw[fermion,thick, >=stealth] (1.5,-.75) to (I2);
    \draw[fermion,thick, >=stealth] (I2) to (1.5,.75);
    \draw[scalarnoarrow, thick] (I1) to (I2);
    \draw (0,0) node[cross out, thick, draw, inner sep=0pt, outer sep=3pt] {} ;
	\node at (-1,0) {$\tau_{2,0}\cdot T~ \;$} ;
 \end{tikzpicture}~~ &= \; a(d,s) \; \; \begin{tikzpicture}[line width=.7 pt, scale=0.6, baseline=(current bounding box.center)
    ]
	\draw (-2,0) to (2,0);
\draw (0,0) circle (1cm);
 \end{tikzpicture} \; + \; b(d,s) \; \; \begin{tikzpicture}[line width=.7 pt, scale=0.6, baseline=(current bounding box.center)
    ]
	\draw (8.8,0) to (9.8,0);
\draw (10.8,0) circle (1cm);
\draw (12.3,1) -- (10.8,1) -- (10.8,-1) -- (12.3,-1);
 \end{tikzpicture} \nonumber \vspace{.3cm} \\
    &=\;a(d,s)\,A_3(d,s)\,+\,b(d,s)\,A_4(d,s)~,
\end{align}
where $a$ and $b$ are rational functions that are in general singular in the limit $d\to4$. From this decomposition it is straightforward to extract two- and three-particle cuts of the corresponding diagram. The former comes in fact entirely from the corresponding cut of $A_4$ on the right-hand side, while the latter is fully determined by the three-particle cut of $A_3$. Schematically, this can be written as
\begin{align}
    \begin{tikzpicture}[line width=.7 pt, scale=0.8, baseline=(current bounding box.center)
    ]
    \draw[fermionbar,thick, >=stealth] (0,0) to (1.5,.75) ;
    \draw[white] (0,0) to coordinate[pos=0.5] (I1) (1.5,-.75) ;
    \draw[fermion,thick, >=stealth] (0,0) to (I1);
    \draw[fermion,thick, >=stealth] (I1) to (1.5,-.75);
    \draw[scalarnoarrow,thick] (1.5,-.75) to (2.5,-.75) ;
    \draw[scalarnoarrow,thick] (1.5,.75) to (2.5,.75) ;
    \draw[white] (1.5,-.75) to coordinate[pos=0.5] (I2) (1.5,.75) ;
    \draw[fermion,thick, >=stealth] (1.5,-.75) to (I2);
    \draw[fermion,thick, >=stealth] (I2) to (1.5,.75);
    \draw[scalarnoarrow, thick] (I1) to (I2);
    \draw (0,0) node[cross out, thick, draw, inner sep=0pt, outer sep=3pt] {} ;
    \draw[dotted] (.5,.8) to (.5,-.8);
 \end{tikzpicture} = b(d,s) \; \; \begin{tikzpicture}[line width=.7 pt, scale=0.6, baseline=(current bounding box.center)
    ]
	\draw (8.8,0) to (9.8,0);
\draw (10.8,0) circle (1cm);
\draw (12.3,1) -- (10.8,1) -- (10.8,-1) -- (12.3,-1);
\draw[dotted] (10.4,1.3) to (10.4,-1.3);
 \end{tikzpicture} ~~~~~,~~~~
    \begin{tikzpicture}[line width=.7 pt, scale=0.8, baseline=(current bounding box.center)
    ]
    \draw[fermionbar,thick, >=stealth] (0,0) to (1.5,.75) ;
    \draw[white] (0,0) to coordinate[pos=0.5] (I1) (1.5,-.75) ;
    \draw[fermion,thick, >=stealth] (0,0) to (I1);
    \draw[fermion,thick, >=stealth] (I1) to (1.5,-.75);
    \draw[scalarnoarrow,thick] (1.5,-.75) to (2.5,-.75) ;
    \draw[scalarnoarrow,thick] (1.5,.75) to (2.5,.75) ;
    \draw[white] (1.5,-.75) to coordinate[pos=0.5] (I2) (1.5,.75) ;
    \draw[fermion,thick, >=stealth] (1.5,-.75) to (I2);
    \draw[fermion,thick, >=stealth] (I2) to (1.5,.75);
    \draw[scalarnoarrow, thick] (I1) to (I2);
    \draw (0,0) node[cross out, thick, draw, inner sep=0pt, outer sep=3pt] {} ;
    \draw[dotted] (1.2,1) to (1.2,-1);
 \end{tikzpicture}
     = \; 
    a(d,s) \; \; \begin{tikzpicture}[line width=.7 pt, scale=0.6, baseline=(current bounding box.center)
    ]
	\draw (-2,0) to (2,0);
\draw (0,0) circle (1cm);
    \draw[dotted] (0,1.3) to (0,-1.3);
 \end{tikzpicture}~~~~.
\end{align}
Evaluating all the two- and three-cut diagrams explicitly with this method we find perfect agreement with the results obtained by other means in Section~\ref{sec:yukawa}.\\

Let us go back to the main object of our study, given in \eq{yukawaff}. In order to have finite correlation functions, we introduce a renormalization constant $Z$ such that
\be
{\bf F} = \frac{{\bf F}_{\rm bare}}{Z}
\ee
becomes finite once ${\bf F}_{\rm bare}$ is expressed as a function of the renormalized couplings. At one-loop order, we find
\be
 {\bf F}^{(1)}= {\bf F}_{0}  \left\{\delta^{(1)} + 
        \left(\frac{y}{4\pi}\right)^2\left[\frac{2}{\epsilon}-2\ln\frac{-s}{\mu^2}+6\right] \right\rbrace,
\ee
where $\delta^{(i)}=(Z^{-1}-1)^{(i)}$, or equivalently $Z=(1+\delta^{(1)}+\delta^{(2)}+\ldots)^{-1}$, and we see that, in order to have a finite form factor, we need to fix
\be
\delta^{(1)}=-\left(\frac{y}{4\pi}\right)^2\frac{2}{\epsilon}\,.
\ee
After adding together the genuine two-loop diagrams and the one-loop counterterm diagrams shown in Fig.~\ref{2loopFeynman}, we are left with a form factor (as a function of the {renormalized} coupling $y$) that reads
\begin{equation} \label{eq:2loop_tmunu}
    \begin{aligned}
        {\bf F}^{(2)}= {\bf F}_{0} \left\{ \delta^{(2)} 
         + \left(\frac{y}{4\pi}\right)^4\left[\,\frac{3}{\epsilon^2}-\frac{5}{2\epsilon}
        -3\ln^2\frac{-s}{\mu^2}+23\ln\frac{-s}{\mu^2}+\text{finite}\,\right]  \right\},
    \end{aligned}
\end{equation}
so that all in all we find
\be
Z^{-1}=1-\left(\frac{y}{4\pi}\right)^2 \frac{2}{\epsilon}-\left(\frac{y}{4\pi}\right)^4\left(\frac{3}{\epsilon^2}-\frac{5}{2\epsilon}\right)+{\cal O}(y^6)\,.
\ee
If we now argue, like we did in the main text, that the divergences found in the form factor of a conserved operator are of infrared nature, we should identify $\gamma_{\rm IR}=-(d\ln Z/d\ln\mu)$, so
\be
\gamma_{\rm IR}=4\left(\frac{y}{4\pi}\right)^2-10\left(\frac{y}{4\pi}\right)^4+{\cal O}(y^6)\,,
\ee
in accordance with the results of Section~\ref{sec:yukawa}. An equivalent way to extract $\gamma_{\rm IR}$ goes through the use of the RG equation, \eq{Ddef}, with $F_{\cal O}\to {\bf F}_0+{\bf F}^{(1)}+{\bf F}^{(2)}+{\cal O}(y^6)$ and $\gamma_{\rm UV}=0$.

\section{Yukawa Fermion Sector} \label{app:fermion_sector}

\subsection{Results for the energy momentum tensor}

In order to perform the same calculation for the form factor of the energy momentum tensor with two external fermions, ${\bf F}^{\mu\nu}(1_{\psi_-},2_{\psi_+})$, the additional 1-loop on-shell amplitudes needed for the 2-cuts are
\begin{equation}
\begin{aligned}
    {\cal A}(1_{\bar{\psi}_-},2_{\bar{\psi}_+} \,;\, 3_{\psi_-}, 4_{\psi_+}) & = \frac{\langle 2 3 \rangle}{\langle 1 4 \rangle} \frac{y^4}{32 \pi^2}  \left[8 \ln \frac{- t}{\mu^2} - 2 \ln \frac{s}{t} - 2 \ln \frac{s}{u} - 16 + \frac{s}{t} \left( \ln^2 \frac{s}{u} + \pi^2 \right) \right], \\
    {\cal A}(1_{\psi_-},2_{\psi_+} \,;\, 3_{\psi_-}, 4_{\psi_+}) & = \frac{[2 4]}{[1 3]} \frac{y^4}{32 \pi^2}  \left[ 2 \ln \frac{s}{u} + 2 \ln \frac{t}{u}  - \left( \ln^2 \frac{s}{t} + \pi^2 \right) \right].
\end{aligned}
\end{equation}
These together with the ones given on the main text lead after summing up all possible 2-cuts to
\be \label{eq:2cut_ferm}
\sum_{\phi_1, \phi_2} {\bf F}^{\mu\nu}(1'_{\phi_1}, 2'_{\phi_2}) S (1'_{\phi_1}, 2'_{\phi_2};\,1_{\psi_-},2_{\psi_+})={\bf F}_0^{\mu\nu}(1_{\psi_-},2_{\psi_+}) \frac{i\pi \,y^4}{(16\pi^2)^2}  \left( 3 \ln\frac{s}{\mu^2} - \frac{51}{2} \right),
\ee
where the sum runs over all non-zero combinations of the fields $\phi_i = {\varphi, \psi, \bar{\psi}}$ and their helicities. \\
For the 3-cuts we also need an additional amplitude which is
\begin{equation}
{\cal A}(1_{\psi_{+}},2_{\psitil_+},3_\varphi\,;\, 4_{\psi_-},5_{\psi_+}) = \frac{y^3}{\langle 1 2 \rangle} \left( \frac{\langle 3 5 \rangle}{\langle 3 4 \rangle} - \frac{[ 3 4 ]}{[ 3 5 ]} \right) + \frac{y^3}{\langle 1 4 \rangle} \left( \frac{\langle 3 5 \rangle}{\langle 2 3 \rangle} + \frac{[ 2 3 ]}{[ 3 5 ]} \right).
\end{equation}
Summing up all possible 3-cuts leads to
\be \label{eq:3cut_ferm}
\sum_{\pm} {\bf F}^{\mu\nu}\!\!\left(1'_{\psi_\pm}, 2'_{\psitil_\pm},3'_\varphi\right) S\left(1'_{\psi_\pm}, 2'_{\psitil_\pm},3'_\varphi\,;\,1_{\psi_-},2_{\psi_+}\right)={\bf F}_0^{\mu\nu}(1_{\psi_-},2_{\psi_+}) \frac{i\pi \,y^4}{(16\pi^2)^2}  \left( 6 \ln\frac{s}{\mu^2} -8 \right).
\ee
Given these results we can proceed as for the case with two external scalars. We take our results \Eq{eq:2cut_ferm} and \Eq{eq:3cut_ferm} and plug them into the master formula \Eq{master2} to find
\begin{equation}
e^{i\pi \left(\gamma_{\rm IR}|_\psi-\frac{2}{y}\,(a_\psi^* +\frac{i\pi}{2}\,\gamma_{\rm IR}|_\psi)\, \beta_y\right)} {\bf F}_0^{\mu\nu}\!\left(1_{\psi_-},2_{\psi_+} \right) = {\bf F}_0^{\mu\nu}(1_{\psi_-},2_{\psi_+}) \frac{i\pi \,y^4}{(16\pi^2)^2} \left(9\ln\frac{s}{\mu^2}-\frac{67}{2}\right).
\end{equation}
After expanding this equation and solving for $\gamma_{\rm IR}$, which by the absence of non-local terms is, as in the scalar case, nothing else than the collinear anomalous dimension $\gamma_{\rm IR} |_\psi = \gamma_{\rm coll} |_\psi$, we find that
\begin{equation}
\gamma^{(1)}_{\rm coll}|_\psi= \left(\frac{\,y^2\,}{16\pi^2}\right) \quad \text{and} \quad \gamma^{(2)}_{\rm coll}|_\psi = - \frac{13}{4} \,\left(\frac{\,y^2\,}{16\pi^2}\right)^2,
\end{equation}
which is, again as for the scalar, nothing else than twice the field anomalous dimensions of $\psi$. \\
We checked the on-shell calculations also by calculating the projected form factor using explicit Feynman integrals. We find
\begin{equation}
    \begin{aligned}
        {\bf F}^{(2)}(1_{\psi_{-}},2_{\psi_{+}})= {\bf F}_{0}(1_{\psi_{-}},2_{\psi_{+}})&\left\{ 1 + \delta^{(1)} + \delta^{(2)} + \frac{1}{2}\left(\frac{y}{4\pi}\right)^2\left[\frac{1}{\epsilon}-\ln\frac{-s}{\mu^2}+3\right] \right. \\
        &\hspace{-3cm} \left.+ \left(\frac{y}{8\pi}\right)^4\left[\frac{18}{\epsilon^2}-\frac{13}{\epsilon}-18\ln^2 \frac{-s}{\mu^2}+134 \ln \frac{-s}{\mu^2}-\frac{519}{2}-6\pi^2\right]+\ldots \right\},
    \end{aligned}
\end{equation}
leading to
\begin{equation}
    Z = \left(1+\delta^{(1)}+\delta^{(2)}+{\cal O}(y^6) \right)^{-1} = \left[1+\left(\frac{y}{4\pi}\right)^2\frac{1}{2\epsilon}+\left(\frac{y}{8\pi}\right)^4\left(\frac{22}{\epsilon^2}-\frac{13}{\epsilon}\right)\right] +{\cal O}(y^6),
\end{equation}
and further to
\be \label{eq:gam_IR_ferm}
\gamma_{\rm IR}= \left(\frac{y}{4\pi}\right)^2-\frac{13}{4}\left(\frac{y}{4\pi}\right)^4+{\cal O}(y^6)\,,
\ee
which confirms the above result.

\subsection{Cross check of the result using the charged current}

A non-trivial cross check of the result presented in this appendix can be obtained by coupling an auxiliary gauge field to our theory. This auxiliary gauge field naturally comes with a conserved current
\begin{equation}
    j^\mu_i=\bar{\psi}_{i}\gamma^\mu\psi_{i}.
\end{equation}
As for the energy momentum tensor, the conservation of the current assures that the form factor of the current, without a propagating gauge boson, does not $\rm UV$ renormalize, hence, the anomalous dimension of the of the current operator has to be equal to $\gamma_{\rm IR}$, as in the case of $T^{\mu \nu}$. \\
In a similar fashion as for the energy momentum tensor we project the spacetime index of the form factor. The current will in general be of the form
\begin{equation}
    j_{\mu}(1_{\psi_{-}},2_{{\psi}_{+}})=\bar{v}(p_2)\mathcal{F}_{\mu}P_Lu(p_1)=\mathbf{F}_{L}(1_{\psi_{-}},2_{{\psi}_{+}})\bar{v}(p_2)\gamma_{\mu}P_Lu(p_1),
\end{equation}
which naturally leads us to define a projection operation by
\begin{equation}
    \mathbf{F}_{L}(1_{\psi_{-}},2_{{\psi}_{+}})=\frac{4}{(D-2)s^2}\text{Tr}\left[\gamma^\mu\slashed{p}_2\mathcal{F}_{\mu}P_L\slashed{p}_1\right],
\end{equation}
to extract the desired form factor $\mathbf{F}_{L}(1_{\psi_{-}},2_{{\psi}_{+}})$. The calculation of $\mathbf{F}_{L}(1_{\psi_{-}},2_{{\psi}_{+}})$ can be performed with either method presented above, on-shell or explicitly using Feynman integrals. The later one directly leads us to
\begin{equation}
    \begin{aligned}
        \mathbf{F}^{(2)}_{L}(1_{\psi_{-}}, & 2_{{\psi}_{+}}) =  \mathbf{F}^{(0)}_{L}(1_{\psi_{-}},2_{{\psi}_{+}}) \left\{ 1 + \delta^{(1)} + \delta^{(2)} + \frac{1}{2}\left(\frac{y}{4\pi}\right)^2\left[\frac{1}{\epsilon}-\ln\frac{-s}{\mu^2}+1\right] \right. \\
        & \left.+ \left(\frac{y}{8\pi}\right)^4\left[\frac{18}{\epsilon^2}-\frac{13}{\epsilon}-18\ln^2\frac{-s}{\mu^2}+62\ln\frac{-s}{\mu^2}-\frac{31}{2}-6\pi^2\right]+\ldots \right\},
    \end{aligned}
\end{equation}
which after determining $Z$ leads exactly to the same result \Eq{eq:gam_IR_ferm}. \\

At this point, one might ask why we used the energy-momentum tensor to calculate the anomalous dimension instead of the vector current. Indeed, using the latter was computationally easier due to the reduced number of diagrams as well as the much simpler vertex rule for the current itself. However, usage of the current is limited to only external states that transform non-trivially under the global symmetry, in this case only the fermions. The energy-momentum tensor, on the other hand, has a non-vanishing form factor with every propagating degree of freedom, i.e., even with those transforming trivially under every internal symmetry of the theory, in this case the scalar field. Further, we can only use the vector current to easily extract the IR anomalous dimension if the associated symmetry is a global one or, equivalently, the corresponding gauge boson is not allowed to propagate in the loops. Otherwise, we can find an identically conserved operator as a counterterm \cite{Collins:2005nj}, such that the UV anomalous dimension is non-vanishing, making it necessary to disentangle UV and IR dynamics. In the end, for a local symmetry, the vector current cannot straightforwardly be used to extract contributions to the IR anomalous dimension proportional to the accompanying gauge coupling. However, these can be obtained using the energy-momentum tensor.


\newpage

\addcontentsline{toc}{section}{Bibliography}

\begin{thebibliography}{99}

\bibitem{Caron-Huot:2016cwu}
S.~Caron-Huot and M.~Wilhelm,
``Renormalization group coefficients and the S-matrix,''
JHEP \textbf{12} (2016), 010
[arXiv:1607.06448 [hep-th]].

\bibitem{EliasMiro:2020tdv}
J.~Elias Mir\'o, J.~Ingoldby and M.~Riembau,
``EFT anomalous dimensions from the S-matrix,''
JHEP \textbf{09} (2020), 163
[arXiv:2005.06983 [hep-ph]].

\bibitem{Baratella:2020lzz}
P.~Baratella, C.~Fernandez and A.~Pomarol,
``Renormalization of Higher-Dimensional Operators from On-shell Amplitudes,''
Nucl. Phys. B \textbf{959} (2020), 115155
[arXiv:2005.07129 [hep-ph]].

\bibitem{EliasMiro:2021jgu}
J.~Elias Miro, C.~Fernandez, M.~A.~Gumus and A.~Pomarol,
``Gearing up for the next generation of LFV experiments, via on-shell methods,''
JHEP \textbf{06} (2022), 126
[arXiv:2112.12131 [hep-ph]].


\bibitem{Bern:2020ikv}
Z.~Bern, J.~Parra-Martinez and E.~Sawyer,
``Structure of two-loop SMEFT anomalous dimensions via on-shell methods,''
JHEP \textbf{10} (2020), 211
doi:10.1007/JHEP10(2020)211
[arXiv:2005.12917 [hep-ph]].

\bibitem{Collins:2005nj}
J.~C.~Collins, A.~V.~Manohar and M.~B.~Wise,
``Renormalization of the vector current in QED,''
Phys. Rev. D \textbf{73} (2006), 105019
[arXiv:hep-th/0512187 [hep-th]].


\bibitem{Baratella:2021guc}
P.~Baratella, D.~Haslehner, M.~Ruhdorfer, J.~Serra and A.~Weiler,
``RG of GR from on-shell amplitudes,''
JHEP \textbf{03} (2022), 156
[arXiv:2109.06191 [hep-th]].

\bibitem{Becher:2009cu}
T.~Becher and M.~Neubert,
``Infrared singularities of scattering amplitudes in perturbative QCD,''
Phys. Rev. Lett. \textbf{102} (2009), 162001
[erratum: Phys. Rev. Lett. \textbf{111} (2013) no.19, 199905]
[arXiv:0901.0722 [hep-ph]].


\bibitem{Becher:2014oda}
T.~Becher, A.~Broggio and A.~Ferroglia,
``Introduction to Soft-Collinear Effective Theory,''
Lect. Notes Phys. \textbf{896} (2015), pp.1-206
[arXiv:1410.1892 [hep-ph]].

\bibitem{Bern:2005iz}
Z.~Bern, L.~J.~Dixon and V.~A.~Smirnov,
``Iteration of planar amplitudes in maximally supersymmetric Yang-Mills theory at three loops and beyond,''
Phys. Rev. D \textbf{72} (2005), 085001
[arXiv:hep-th/0505205 [hep-th]].



\bibitem{Baratella:2020dvw}
P.~Baratella, C.~Fernandez, B.~von Harling and A.~Pomarol,
``Anomalous Dimensions of Effective Theories from Partial Waves,''
JHEP \textbf{03} (2021), 287
[arXiv:2010.13809 [hep-ph]].

\bibitem{Mastrolia:2009dr}
P.~Mastrolia,
``Double-Cut of Scattering Amplitudes and Stokes' Theorem,''
Phys. Lett. B \textbf{678} (2009), 246-249
[arXiv:0905.2909 [hep-ph]].

\bibitem{Arkani-Hamed:2008owk}
N.~Arkani-Hamed, F.~Cachazo and J.~Kaplan,
``What is the Simplest Quantum Field Theory?,''
JHEP \textbf{09} (2010), 016
[arXiv:0808.1446 [hep-th]].

\bibitem{Yang:2019vag}
G.~Yang,
``On-shell methods for form factors in $\mathcal{N}=4$ SYM and their applications,''
Sci. China Phys. Mech. Astron. \textbf{63} (2020) no.7, 270001
[arXiv:1912.11454 [hep-th]].

\bibitem{Machacek:1983tz}
M.~E.~Machacek and M.~T.~Vaughn,
Nucl. Phys. B \textbf{222} (1983), 83-103
doi:10.1016/0550-3213(83)90610-7

\bibitem{Cheung:2017pzi}
C.~Cheung,
``TASI Lectures on Scattering Amplitudes,''
[arXiv:1708.03872 [hep-ph]].

\bibitem{Arkani-Hamed:2017jhn}
N.~Arkani-Hamed, T.~C.~Huang and Y.~t.~Huang,
``Scattering amplitudes for all masses and spins,''
JHEP \textbf{11} (2021), 070
[arXiv:1709.04891 [hep-th]].

\bibitem{Jiang:2020mhe}
M.~Jiang, T.~Ma and J.~Shu,
``Renormalization Group Evolution from On-shell SMEFT,''
JHEP \textbf{01} (2021), 101
[arXiv:2005.10261 [hep-ph]].


\bibitem{Shadmi:2018xan}
Y.~Shadmi and Y.~Weiss,
``Effective Field Theory Amplitudes the On-Shell Way: Scalar and Vector Couplings to Gluons,''
JHEP \textbf{02} (2019), 165
[arXiv:1809.09644 [hep-ph]].

\bibitem{Ma:2019gtx}
T.~Ma, J.~Shu and M.~L.~Xiao,
``Standard Model Effective Field Theory from On-shell Amplitudes,''
[arXiv:1902.06752 [hep-ph]].

\bibitem{Aoude:2019tzn}
R.~Aoude and C.~S.~Machado,
``The Rise of SMEFT On-shell Amplitudes,''
JHEP \textbf{12} (2019), 058
[arXiv:1905.11433 [hep-ph]].

\bibitem{Durieux:2019siw}
G.~Durieux and C.~S.~Machado,
``Enumerating higher-dimensional operators with on-shell amplitudes,''
Phys. Rev. D \textbf{101} (2020) no.9, 095021
[arXiv:1912.08827 [hep-ph]].

\bibitem{AccettulliHuber:2021uoa}
M.~Accettulli Huber and S.~De Angelis,
``Standard Model EFTs via on-shell methods,''
JHEP \textbf{11} (2021), 221
[arXiv:2108.03669 [hep-th]].


\bibitem{Cheung:2015aba}
C.~Cheung and C.~H.~Shen,
``Nonrenormalization Theorems without Supersymmetry,''
Phys. Rev. Lett. \textbf{115} (2015) no.7, 071601
[arXiv:1505.01844 [hep-ph]].

\bibitem{Craig:2019wmo}
N.~Craig, M.~Jiang, Y.~Y.~Li and D.~Sutherland,
``Loops and Trees in Generic EFTs,''
JHEP \textbf{08} (2020), 086
[arXiv:2001.00017 [hep-ph]].

\bibitem{Jiang:2020rwz}
M.~Jiang, J.~Shu, M.~L.~Xiao and Y.~H.~Zheng,
``Partial Wave Amplitude Basis and Selection Rules in Effective Field Theories,''
Phys. Rev. Lett. \textbf{126} (2021) no.1, 011601
[arXiv:2001.04481 [hep-ph]].

\bibitem{Maierhofer:2017gsa}
P.~Maierh\"ofer, J.~Usovitsch and P.~Uwer,
Comput. Phys. Commun. \textbf{230} (2018), 99-112
doi:10.1016/j.cpc.2018.04.012
[arXiv:1705.05610 [hep-ph]].

\bibitem{Gehrmann:1999as}
T.~Gehrmann and E.~Remiddi,
Nucl. Phys. B \textbf{580} (2000), 485-518
doi:10.1016/S0550-3213(00)00223-6
[arXiv:hep-ph/9912329 [hep-ph]].

\bibitem{Gehrmann:2005pd}
T.~Gehrmann, T.~Huber and D.~Maitre,
Phys. Lett. B \textbf{622} (2005), 295-302
doi:10.1016/j.physletb.2005.07.019
[arXiv:hep-ph/0507061 [hep-ph]].


\end{thebibliography}
\end{document}